\documentclass[10pt,journal,compsoc]{IEEEtran}
\ifCLASSOPTIONcompsoc
  \usepackage[nocompress]{cite}
\else
  \usepackage{cite}
\fi
\usepackage[pdftex]{graphicx}
\usepackage{url}
\usepackage[utf8]{inputenc}
\usepackage{xcolor}
\usepackage{amsmath}
\usepackage{amssymb}
\usepackage{xifthen}
\usepackage[nolist,nohyperlinks]{acronym}
\usepackage[ruled,vlined]{algorithm2e}
\usepackage[hidelinks]{hyperref}
\usepackage[capitalise]{cleveref}
\usepackage{bbm}
\usepackage{numprint}

\begin{acronym}
\acro{CWS}{consistent weighted sampling}
\acro{0-bit CWS}{0-bit consistent weighted sampling}
\acro{ICWS}{improved consistent weighted sampling}
\acro{PRNG}{pseudorandom number generator}
\acro{MSE}{mean squared error}
\acro{OPH}{One Permutation Hashing}
\acro{FSS}{Fast Similarity Sketching}
\end{acronym}

\newcommand*{\symDefine}[2]{\newcommand{{#1}}{{#2}}}
\newcommand\symVariance{\operatorname{Var}}
\newcommand\symExpectation{\operatorname{\mathbb{E}}}

\symDefine{\symBoundary}{a}
\symDefine{\symSetA}{A}
\symDefine{\symArea}{A}
\symDefine{\symBBit}{b}
\symDefine{\symSetB}{B}
\symDefine{\symBuffer}{b}
\symDefine{\symConstant}{c}
\symDefine{\symBigConstant}{C}
\symDefine{\symNumExamples}{c}
\symDefine{\symVersionCounter}{c}
\symDefine{\symInputElement}{d}
\symDefine{\symUniverse}{D}
\symDefine{\symFunction}{f}
\symDefine{\symDistribution}{F}
\symDefine{\symPermutationArray}{g}
\symDefine{\symHashValue}{h}
\symDefine{\symHarmonic}{H}
\symDefine{\symIndex}{i}
\symDefine{\symMultiSet}{I}
\symDefine{\symJaccard}{J}
\symDefine{\symJaccardW}{\symJaccard_W}
\symDefine{\symJaccardN}{\symJaccard_N}
\symDefine{\symJaccardP}{\symJaccard_P}
\symDefine{\symJaccardEstimator}{\hat{J}}
\symDefine{\symIndexJ}{j}
\symDefine{\symHashIndex}{k}
\symDefine{\symIndexL}{l}
\symDefine{\symLabel}{l}
\symDefine{\symSequenceIndices}{L}
\symDefine{\symHashSize}{m}
\symDefine{\symInputSize}{n}
\symDefine{\symBigO}{\mathcal{O}}
\symDefine{\symFirstOccurP}{p}
\symDefine{\symProbability}{\Pr}
\symDefine{\symHashMin}{q}
\symDefine{\symMultiplicity}{r}
\symDefine{\symRNG}{R}
\symDefine{\symSetS}{S}
\symDefine{\symBufferCounter}{s}
\symDefine{\symBufferCounterTwo}{t}
\symDefine{\symUniformHash}{u}
\symDefine{\symUnionSize}{u}
\symDefine{\symVersionArray}{v}
\symDefine{\symWeight}{w}
\symDefine{\symWeightInverse}{\symWeight_\text{inv}}
\symDefine{\symTestWeightSet}{W}
\symDefine{\symPoint}{x}
\symDefine{\symSequencePoint}{X}
\symDefine{\symY}{y}
\symDefine{\symSignature}{z}
\symDefine{\symZScore}{z}
\symDefine{\symSignatureReduced}{\hat{\symSignature}}
\symDefine{\symHashValueMax}{\symHashMin_\text{max}}
\symDefine{\symPermutation}{\pi}
\symDefine{\symImprovement}{\alpha}
\symDefine{\symRate}{\lambda}
\symDefine{\symIndicator}{\mathbbm{1}}
\symDefine{\symDensity}{\rho}
\symDefine{\symOmega}{\omega}
\symDefine{\symTailIndex}{\eta}
\symDefine{\symXor}{\text{XOR}}
\symDefine{\symBeta}{\beta}
\symDefine{\symGamma}{\gamma}
\symDefine{\symDelta}{\delta}
\symDefine{\symMu}{\mu}

\DeclareMathOperator*{\argmin}{arg\,min}

\DeclareMathOperator*{\symExponential}{Exp}
\DeclareMathOperator*{\symUniform}{Uniform}
\DeclareMathOperator*{\symPareto}{Pareto}

\newcommand{\myAlg}[2][]{
    \ifthenelse{\isempty{#1}}%
    {\begin{figure}}
    {\begin{figure}[#1]}
    \begingroup 
    \csname @twocolumnfalse\endcsname
    \noindent
    \resizebox{\columnwidth}{!}{%
        \begin{minipage}{1.4\columnwidth}
            \begin{algorithm}[H]
                \DontPrintSemicolon
                {#2}
            \end{algorithm}
        \end{minipage}%
    }
    \endgroup
    \end{figure}
}

\SetKwComment{Com}{}{}
\SetKwFunction{InitPermutationGenerator}{InitPermutationGenerator}
\SetKwFunction{GenerateNextPermutationElement}{GenerateNextPermutationElement}
\SetKwFunction{ResetPermutationGenerator}{ResetPermutationGenerator}
\SetKwProg{Function}{def}{}{}
\SetKw{Break}{break}
\SetKw{Continue}{continue}
\SetKwBlock{Loop}{loop}{end}
\SetKw{New}{new}
\SetKw{And}{and} 
\SetKw{Not}{not}
\SetArgSty{textnormal}
\SetFuncSty{textnormal}

\SetCommentSty{mycommfont}

\begin{document}

\title{ProbMinHash -- A Class of Locality-Sensitive Hash Algorithms for the (Probability) \\ Jaccard Similarity}
\author{Otmar~Ertl
\IEEEcompsocitemizethanks{\IEEEcompsocthanksitem O. Ertl is affiliated with Dynatrace Research, Linz, Austria.\protect\\
E-mail: otmar.ertl@dynatrace.com or otmar.ertl@gmail.com
}
\thanks{Manuscript received XXXX XX, 2020; revised XXXX XX, XXXX.\newline
The final version is available at \newline
\url{http://dx.doi.org/10.1109/TKDE.2020.3021176}.}}
\markboth{IEEE Transactions on Knowledge and Data Engineering,~Vol.~XX, No.~X, XXXX~2020}%
{Ertl: ProbMinHash -- A Class of Locality-Sensitive Hash Algorithms for the (Probability) Jaccard Similarity}

\IEEEpubid{Copyright (c) 2020 IEEE. Personal use is permitted. For any other purposes, permission must be obtained from the IEEE by emailing pubs-permissions@ieee.org.}


\IEEEtitleabstractindextext{%
\begin{abstract}
The probability Jaccard similarity was recently proposed as a natural generalization of the Jaccard similarity to measure the proximity of sets whose elements are associated with relative frequencies or probabilities. In combination with a hash algorithm that maps those weighted sets to compact signatures which allow fast estimation of pairwise similarities, it constitutes a valuable method for big data applications such as near-duplicate detection, nearest neighbor search, or clustering. This paper introduces a class of one-pass locality-sensitive hash algorithms that are orders of magnitude faster than the original approach. The performance gain is achieved by calculating signature components not independently, but collectively. Four different algorithms are proposed based on this idea. Two of them are statistically equivalent to the original approach and can be used as drop-in replacements. The other two may even improve the estimation error by introducing statistical dependence between signature components. Moreover, the presented techniques can be specialized for the conventional Jaccard similarity, resulting in highly efficient algorithms that outperform traditional minwise hashing and that are able to compete with the state of the art.
\end{abstract}

}

\maketitle
\IEEEdisplaynontitleabstractindextext
\IEEEpeerreviewmaketitle
\IEEEraisesectionheading{\section{Introduction}\label{sec:introduction}}
\IEEEPARstart{T}{he} calculation of pairwise object similarities is an important task for clustering, near-duplicate detection, or nearest neighbor search. Big data applications require sophisticated algorithms to overcome time and space constraints. A widely used technique to reduce costs for pairwise similarity computations is minwise hashing (MinHash) \cite{Broder1997}. 
It allows the calculation of signatures for individual objects that can be represented as sets of features. Using only the corresponding signatures, the Jaccard similarity can be estimated from the number of equal signature components.

The Jaccard similarity $\symJaccard$ of two sets $\symSetA$ and $\symSetB$ is given by 
\begin{equation*} 
\symJaccard
= 
\frac{
|\symSetA\cap \symSetB|
}{
|\symSetA\cup \symSetB|
}.
\end{equation*}
Minwise hashing maps a set $\symSetS$ to an $\symHashSize$-dimensional signature vector $\symSignature(\symSetS) = (\symSignature_1(\symSetS),\symSignature_2(\symSetS),\ldots,\symSignature_{\symHashSize}(\symSetS))$ with statistically independent components that are defined as 
\begin{equation}
\label{equ:signature_def}
\symSignature_{\symHashIndex}(\symSetS)
:=
\argmin_{\symInputElement\in\symSetS}
\symHashValue_\symHashIndex(\symInputElement)
\end{equation}
where $\symHashValue_\symHashIndex$ are independent hash functions with identically distributed output. If hash collisions of $\symHashValue_\symHashIndex$ are very unlikely and can be ignored, the probability that the same signature components of two different sets $\symSetA$ and $\symSetB$ have the identical value is equal to the Jaccard similarity
\begin{equation}
\label{equ:equal_probability}
\symProbability(\symSignature_{\symHashIndex}(\symSetA)
=
\symSignature_{\symHashIndex}(\symSetB))
=
\symJaccard.
\end{equation}
This property allows unbiased estimation of $\symJaccard$ using the estimator
\begin{equation}
\label{equ:jaccard_estimator}
\symJaccardEstimator
(\symSignature(\symSetA),\symSignature(\symSetB))
=
\frac{1}{\symHashSize}
\sum_{\symHashIndex=1}^\symHashSize \symIndicator(
\symSignature_{\symHashIndex}(\symSetA)
=
\symSignature_{\symHashIndex}(\symSetB))
\end{equation}
where $\symIndicator$ denotes the indicator function. The variance of this estimator is
\begin{equation}
\label{equ:jaccard_variance}
\symVariance(\symJaccardEstimator
(\symSignature(\symSetA),\symSignature(\symSetB))
)
=
\frac{\symJaccard (1-\symJaccard)}{\symHashSize},
\end{equation}
because the number of equal components in $\symSignature(\symSetA)$ and $\symSignature(\symSetB)$ is binomially distributed with success probability $\symJaccard$, if the signature components are independent. 

\subsection{Incorporating Weights}
Describing objects as feature sets is not always appropriate, because the features are sometimes associated with weights. For example, text documents can be represented as bag of words weighted according to their term frequency–inverse document frequency \cite{Leskovec2014}.
For the mathematical representation of weighted sets we use weight functions which map elements to their associated nonnegative weights. Weight functions can be treated as a mapping from the universe of all possible elements $\symUniverse$, if elements not belonging to the corresponding weighted set are considered to have a weight equal to 0.

Multiple approaches have been proposed to generalize the Jaccard similarity to weighted sets.  
The weighted Jaccard similarity $\symJaccardW$ is one possibility to generalize the Jaccard similarity $\symJaccard$ and is defined as
\begin{equation}
\label{equ:def_weighted_jaccard}
\symJaccardW = \frac{
\sum_{\symInputElement\in\symUniverse}\min(\symWeight_\symSetA(\symInputElement),\symWeight_\symSetB(\symInputElement))
}{
\sum_{\symInputElement\in\symUniverse}\max(\symWeight_\symSetA(\symInputElement),\symWeight_\symSetB(\symInputElement))
}.
\end{equation}
Here $\symWeight_\symSetA$ and $\symWeight_\symSetB$ are the weight functions for weighted sets $\symSetA$ and $\symSetB$, respectively.
$\symJaccard$ and $\symJaccardW$ are identical for a set whose elements have a weight of $1$ while all other elements of the universe $\symUniverse$ have a weight of $0$. Various hash algorithms have been developed to calculate signatures that allow the estimation of $\symJaccardW$ using estimator \eqref{equ:jaccard_estimator} \cite{Ertl2018, Ioffe2010, Shrivastava2016}.   

Sometimes the scale of feature weights is not important. This is for example the case if the weights represent relative frequencies or probabilities. Multiplying all weights of a set by the same factor should not change the similarities to other sets. An obvious solution to achieve this scale-invariance is the normalization of weights before calculating $\symJaccardW$ \cite{Li2015, Yang2017}. This leads to the normalized weighted Jaccard similarity
\begin{equation*}
\symJaccardN = \frac{
\sum_{\symInputElement\in\symUniverse}
\min\!
\left(
\frac{\symWeight_\symSetA(\symInputElement)}{\sum_{\symInputElement'\in\symUniverse}\symWeight_\symSetA(\symInputElement')}
,
\frac{\symWeight_\symSetB(\symInputElement)}{\sum_{\symInputElement'\in\symUniverse}\symWeight_\symSetB(\symInputElement')}
\right)
}{
\sum_{\symInputElement\in\symUniverse}
\max\!
\left(
\frac{\symWeight_\symSetA(\symInputElement)}{\sum_{\symInputElement'\in\symUniverse}\symWeight_\symSetA(\symInputElement')}
,
\frac{\symWeight_\symSetB(\symInputElement)}{\sum_{\symInputElement'\in\symUniverse}\symWeight_\symSetB(\symInputElement')}
\right)
}.
\end{equation*}
In this way the same locality-sensitive hash algorithms can be used as for $\symJaccardW$.

\myAlg{
\caption{P-MinHash.}
\label{alg:p_min_hash}
\KwIn{$\symWeight$}
\KwOut{$\symSignature_1,
\symSignature_2,
\ldots,
\symSignature_{\symHashSize}$}
$(
\symHashMin_1,
\symHashMin_2,
\ldots,
\symHashMin_{\symHashSize}
)\gets
(\infty,\infty,\ldots,\infty)$\;
\ForAll{$\symInputElement\in\symUniverse$ such that $\symWeight(\symInputElement) > 0$}
{
$\symWeightInverse\gets 1/\symWeight(\symInputElement)$\;
$\symRNG \gets$ \New \acs{PRNG} with seed $\symInputElement$\;
\For{$\symHashIndex \gets 1$ \KwTo $\symHashSize$}{
$\symHashValue\gets \symWeightInverse\cdot\symRNG[\symExponential(1)]$\;
\If{$\symHashValue < \symHashMin_\symHashIndex$}
{
    $\symHashMin_\symHashIndex \gets \symHashValue$\;
    $\symSignature_\symHashIndex \gets \symInputElement$\;
}
}
}
}

Recently, another scale-invariant metric was proposed called probability Jaccard similarity
\begin{equation*}
\symJaccardP =
\sum_{\symInputElement\in\symUniverse}
\frac{1}{
\sum_{\symInputElement'\in\symUniverse}
\max\!\left(
\frac{
\symWeight_\symSetA(\symInputElement')
}{
\symWeight_\symSetA(\symInputElement)
},
\frac{
\symWeight_\symSetB(\symInputElement')
}{
\symWeight_\symSetB(\symInputElement)
}
\right)
}.
\end{equation*}
Since $\symJaccardP$  is Pareto optimal \cite{Moulton2018}, which is a property not shared by $\symJaccardN$, it is a more natural extension of $\symJaccard$ to discrete probability distributions. The probability Jaccard similarity $\symJaccardP$ was analyzed in detail and presented together with an appropriate locality-sensitive hash algorithm in \cite{Moulton2018}. 
The algorithm uses the hash functions
\begin{equation}
\label{equ:def_hash_jp}
\symHashValue_\symHashIndex(\symInputElement)
\sim
\symExponential(\symRate\symWeight(\symInputElement))
\sim
\frac{1}{\symRate\symWeight(\symInputElement)}\symExponential(1),
\end{equation}
which yield hash values distributed exponentially with rate proportional to the weight $\symWeight(\symInputElement)$ of the given element $\symInputElement$. The proportionality constant $\symRate$ is a free parameter and has no influence on the signature defined by \eqref{equ:signature_def}. It was shown \cite{Moulton2018} that the resulting signature satisfies
$\symProbability(\symSignature_{\symHashIndex}(\symSetA)
=
\symSignature_{\symHashIndex}(\symSetB))
=
\symJaccardP$ and therefore allows unbiased estimation of $\symJaccardP$ from the proportion of equal components using estimator \eqref{equ:jaccard_estimator}.
If the hash values $\symHashValue_{\symHashIndex}(\symInputElement)$ with $\symHashIndex\in\{1,2,\ldots,\symHashSize\}$ are statistically independent, the signature components will be independent as well and the variance of the estimator will be $\symJaccardP(1-\symJaccardP)/\symHashSize$ analogous to \eqref{equ:jaccard_variance}.
As $\symJaccardW$, $\symJaccardP$ corresponds to $\symJaccard$ in case of binary weights $\symWeight(\symInputElement)\in\{0, 1\}$ and can therefore be regarded as generalization of the Jaccard similarity $\symJaccard$. The P-MinHash signature is equivalent to the MinHash signature \eqref{equ:signature_def} in this case, because the hash functions $\symHashValue_\symHashIndex$ defined by \eqref{equ:def_hash_jp} are identically distributed, if $\symWeight(\symInputElement)$ is always equal to 1.

The signature definition \eqref{equ:signature_def} together with hash functions \eqref{equ:def_hash_jp} can be straightforwardly translated into an algorithm called P-MinHash \cite{Moulton2018} shown as \cref{alg:p_min_hash}. Instead of using $\symHashSize$ independent hash functions, we use a \ac{PRNG} $\symRNG$ seeded with a hash value of $\symInputElement$ to generate independent and exponentially distributed values. Since the rate parameter of the exponential distributed random values in \eqref{equ:def_hash_jp} is a free parameter, $\symRate=1$ is used for simplicity. $\symRNG[\symExponential(\symRate)]$ denotes the generation of an exponentially distributed random value with rate parameter $\symRate$ using random bits taken from $\symRNG$. Since floating-point divisions are more expensive than multiplications, it makes sense to precalculate the reciprocal weight $\symWeightInverse$ as done in \cref{alg:p_min_hash}.

\subsection{Related Work}

Interestingly, hash algorithms with collision probabilities equal to $\symJaccardP$ have already been unintentionally presented before $\symJaccardP$ was actually discovered and thoroughly analyzed in \cite{Moulton2018}. In \cite{Yang2017} a data structure called HistoSketch was proposed to calculate signatures for $\symJaccardN$. The derivation started from \ac{0-bit CWS} \cite{Li2015}, which was proposed as simplification of \ac{ICWS} \cite{Ioffe2010}. While the collision probability for \ac{ICWS} equals $\symJaccardW$, it is actually not known for \ac{0-bit CWS} and may be far off from $\symJaccardW$ \cite{Ertl2018}. For example, consider two weighted sets, both consisting of the same single element. \ac{0-bit CWS} will always have a collision probability of 1 regardless of the actual weights which contradicts \eqref{equ:def_weighted_jaccard}. Therefore, choosing this algorithm with undefined behavior as starting point for the derivation of a new algorithm is questionable. 
Nevertheless, after some simplifications and thanks to a nonequivalent transformation that eliminated the scale dependence, the final HistoSketch algorithm had a collision probability equal to $\symJaccardP$ instead of the originally desired $\symJaccardN$. Without awareness of the correct collision probability, the same authors simplified the HistoSketch algorithm \cite{Yang2019a} and finally obtained an algorithm that was equivalent to P-MinHash \cite{Moulton2018}.
 
Another but similar attempt to derive a simplified algorithm from \ac{0-bit CWS} is given in \cite{Raff2018}. However, the use of a slightly different nonequivalent transformation also led to a slightly different hash algorithm. Even though the algorithm is scale-invariant, the hash collision probability is neither equal to $\symJaccardN$ nor equal to $\symJaccardP$. 

The straightforward implementation of signatures based on \eqref{equ:signature_def} leads to time complexities of $\symBigO(\symInputSize\symHashSize)$, where $\symHashSize$ denotes the signature size and $\symInputSize$ is the set size or the number of elements with nonzero weight $\symInputSize = |\{ \symInputElement : \symWeight(\symInputElement) > 0\}|$ in the weighted case. A lot of effort was done to break this $\symBigO(\symInputSize\symHashSize)$ barrier.
By calculating all $\symHashSize$ signature components in a more collective fashion, much better time complexities of kind $\symBigO(\symInputSize + \symFunction(\symHashSize))$ are possible, where $\symFunction$ is some function only dependent on $\symHashSize$. In case of the conventional Jaccard similarity $\symJaccard$,
\ac{OPH} \cite{Li2012,Mai2019,Shrivastava2017,Shrivastava2014a}, \ac{FSS} \cite{Dahlgaard2017}, or the SuperMinHash algorithm \cite{Ertl2017} are representatives of such algorithms. They have in common that signature components are not statistically independent. Because of that, the latter two algorithms even have the property that estimation errors are significantly reduced, if $\symInputSize$ is not much larger than $\symHashSize$.

\begin{figure}
    \centering
    \includegraphics[width=0.6\columnwidth]{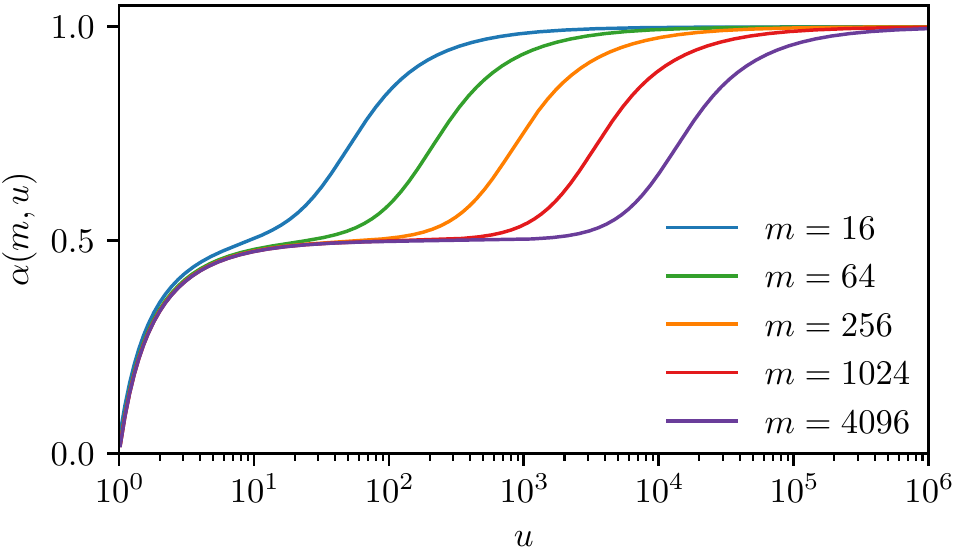}
    \caption{The function $\symImprovement(\symHashSize,\symUnionSize)$ over $\symUnionSize:=|\symSetA\cup\symSetB|$ for different values of $\symHashSize$.}
    \label{fig:alpha}
\end{figure}
    
As example, the SuperMinHash algorithm \cite{Ertl2017} defines the hash functions $\symHashValue_\symHashIndex$ as
\begin{equation*}
\symHashValue_\symHashIndex(\symInputElement)
:=
\symUniformHash_{\symHashIndex}(\symInputElement) + \symPermutation_{\symHashIndex}(\symInputElement).
\end{equation*}
Here $\symUniformHash_{\symHashIndex}$ is uniformly distributed over $[0,1)$ and $\symPermutation_{\symHashIndex}(\symInputElement)$ are the elements of a random permutation
of values $\{0,1,2,\ldots,\symHashSize-1\}$.
The corresponding signature given by \eqref{equ:signature_def}
satisfies \eqref{equ:equal_probability} and the variance of the estimator \eqref{equ:jaccard_estimator} is
\begin{equation}
\label{equ:superminhash_variance}
\symVariance(\symJaccardEstimator)
=
\frac{\symJaccard(1-\symJaccard)}{\symHashSize}
\symImprovement(\symHashSize,\symUnionSize)
\end{equation}
where $\symUnionSize := |\symSetA \cup \symSetB|$ denotes the union cardinality. $\symImprovement(\symHashSize,\symUnionSize)$ is given by 
\begin{equation}
\label{equ:improvement_factor}
\symImprovement(\symHashSize,\symUnionSize)
:=
1
-
\frac{\sum_{\symIndexL=1}^{\symHashSize-1}
\symIndexL^{\symUnionSize}
\left(
(\symIndexL + 1)^{\symUnionSize}
+
(\symIndexL - 1)^{\symUnionSize}
-
2
{\symIndexL}^{\symUnionSize}
\right)}{(\symHashSize-1)^{\symUnionSize-1} \symHashSize^{\symUnionSize} (\symUnionSize-1)}
\end{equation}
and shown in \cref{fig:alpha} for different values of $\symHashSize$. For $\symUnionSize<\symHashSize$ the function value tends to be in the range of $0.5$, and therefore, the variance \eqref{equ:superminhash_variance} is significantly smaller than that of the original MinHash algorithm given by \eqref{equ:jaccard_variance}.

The first algorithm for the conventional Jaccard similarity $\symJaccard$ that has overcome the $\symBigO(\symInputSize\symHashSize)$ barrier with provable statistically independent signature components was presented in \cite{Ertl2018} as a special case of the BagMinHash algorithm. More generally, BagMinHash is the first algorithm that has overcome the $\symBigO(\symInputSize\symHashSize)$ barrier for $\symJaccardW$. Only for special cases with beforehand known universe and upper bounds for weights another fast approach was presented before in \cite{Shrivastava2016}.

A time complexity proportional to $\symInputSize\symHashSize$ would not be a big problem, if both $\symInputSize$ and $\symHashSize$ were small. However, real world problems often have large feature sizes $\symInputSize$. Moreover, it is common that the signature size $\symHashSize$ is in the hundreds or even thousands \cite{Ioffe2010, Li2015, Nissim2019, Raff2019, Rowe2019, Shrivastava2014a}. 
In particular, indexing techniques like locality-sensitive hashing \cite{Bawa2005, Indyk1998, Lv2007}, which enable sublinear nearest neighbor lookups, require many signature components to increase sensitivity and specificity.

Even larger signature sizes are needed, if $\symBBit$-bit minwise hashing is used \cite{Li2010}. This technique reduces each signature component to only a few bits. The loss of information must be compensated by increasing the number of components in order to achieve the same estimation error. Nevertheless, this approach can significantly reduce the total space requirements of signatures, especially if one is mainly interested in high similarities. 
Any signature can be easily reduced to a $\symBBit$-bit signature using \cref{alg:bbit} which transforms each component by taking $\symBBit$ bits from a hash value calculated from the component itself and its index. Since $\symBBit$-bit integer values of different elements will collide with high probability, estimator \eqref{equ:jaccard_estimator} will have a bias that must be accounted for. Moreover, to avoid correlated collisions of different elements over multiple signature components, it is crucial that the involved hash value computation also includes the component index as done by \cref{alg:bbit}.

The cosine similarity is a widely-used alternative to the Jaccard similarity metrics for weighted sets for which also well established locality-sensitive hash algorithms exist \cite{Charikar2002, Andoni2015}. Even though they can be efficiently implemented, they also do not overcome the $\symBigO(\symInputSize\symHashSize)$ time complexity. A hash algorithm designed for a certain metric is not necessarily the best option to be used in the context of locality-sensitive hashing \cite{Christiani2017}. For example, $\symBBit$-bit minwise hashing outperforms SimHash even for the cosine similarity in case of binary weights \cite{Shrivastava2014}. Fast hash algorithms with time complexities below $\symBigO(\symInputSize\symHashSize)$ could therefore also be valuable for other metrics.

\subsection{Applications}
The probability Jaccard similarity $\symJaccardP$ is a metric that was discovered only recently \cite{Moulton2018} and is therefore not yet very widespread. However, since HistoSketch \cite{Yang2017, Yang2019a} finally turned out to be a hash algorithm for $\symJaccardP$, it has already been successfully used to calculate graph embeddings \cite{Yang2019} or to create sketches of k-mer spectra for microbiome analytics \cite{Rowe2019}. A faster hash algorithm with a time complexity below $\symBigO(\symInputSize\symHashSize)$ could directly improve the performance of those applications, making them even more attractive.
At the same time, a fast algorithm would make $\symJaccardP$ more appealing for many more applications for which other metrics such as the Jaccard or cosine similarity were previously preferred because of their well-known hash algorithms.

\subsection{Our Contributions}

\myAlg{
\caption{Reduction to a $\symBBit$-bit signature.}
\label{alg:bbit}
\KwIn{$\symSignature_1,
\symSignature_2,
\ldots,
\symSignature_{\symHashSize}$}
\KwOut{$\symSignatureReduced_1,
\symSignatureReduced_2,
\ldots,
\symSignatureReduced_{\symHashSize}\in\{0,1,2,\ldots,2^{\symBBit}-1\}$}
\For{$\symIndex\gets 1,2,\ldots,\symHashSize$}{
$\symSignatureReduced_\symIndex \gets$ least significant $\symBBit$ bits of hash value calculated from pair $(\symSignature_\symIndex,\symIndex)$ \;
}
}

Motivated by recently developed algorithms like SuperMinHash \cite{Ertl2017} for the conventional Jaccard similarity $\symJaccard$  or BagMinHash \cite{Ertl2018} for the weighted Jaccard similarity $\symJaccardW$, which both achieved superior performance by calculating signature components in a collective fashion instead of calculating them independently, we applied the same principle to design new minwise-hashing algorithms for the probability Jaccard similarity $\symJaccardP$ \cite{Moulton2018}. These algorithms are the first to undercut the $\symBigO(\symInputSize\symHashSize)$ runtime complexity of the state-of-the-art P-MinHash algorithm. 

We present four different ways to generate exponentially distributed hash values satisfying \eqref{equ:def_hash_jp}, which then can be directly translated into four new one-pass algorithms called ProbMinHash1, ProbMinHash2, ProbMinHash3, and ProbMinHash4, respectively. All of them are, due to the much better time complexity, orders of magnitude faster than the original P-MinHash algorithm with the exception of very small input sizes $\symInputSize$. The first two are statistically equivalent to the original approach. The latter two are even able to reduce the estimation error due to the statistical dependence of individual signature components, as our experimental results will show. Similar to the SuperMinHash algorithm the variance of estimator \eqref{equ:jaccard_estimator} is decreased by up to a factor of two for input sizes $\symInputSize$ smaller than the signature size $\symHashSize$.

\begin{figure*}
\centering
\includegraphics[width=1\textwidth]{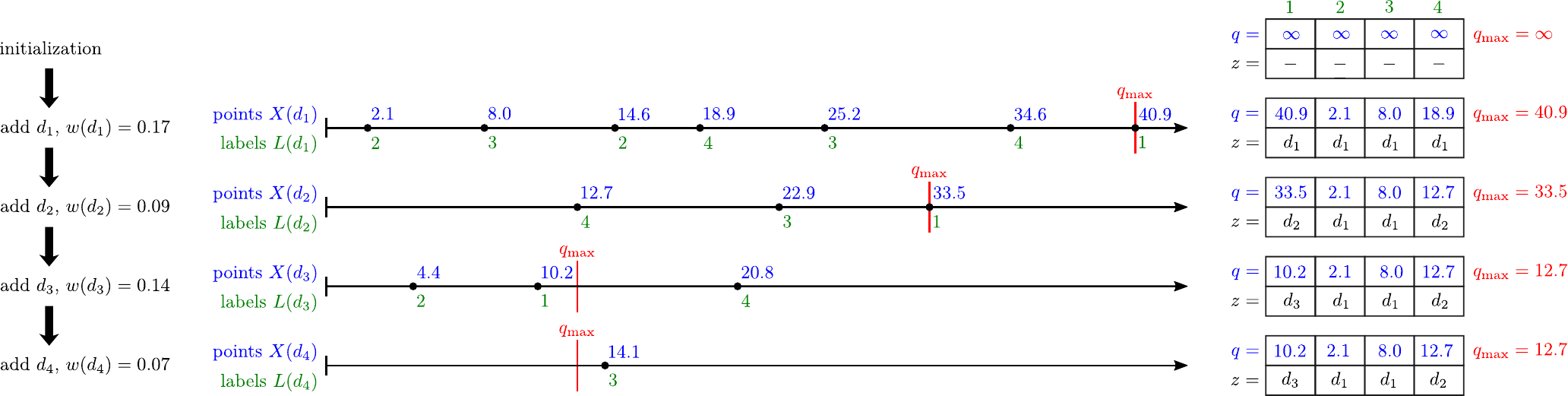}
\caption{Illustration of \cref{alg:prob_min_hash} for $\symHashSize=4$ and a set of 4 weighted elements $\{ \symInputElement_1, \symInputElement_2, \symInputElement_3, \symInputElement_4 \}$. For each element $\symInputElement$ a random sequence of points $\symSequencePoint(\symInputElement)$ and a random sequence of labels $\symSequenceIndices(\symInputElement)$ are generated until the stop limit $\symHashValueMax$ is reached or exceeded. The density of points is proportional to the corresponding weight $\symWeight(\symInputElement)$. If a generated point $\symPoint_\symIndex(\symInputElement)$ with label $\symLabel_\symIndex(\symInputElement)$ is smaller than $\symHashMin_{\symLabel_\symIndex(\symInputElement)}$ its value gets replaced. At the same time the corresponding $\symSignature_{\symLabel_\symIndex(\symInputElement)}$ is updated with $\symInputElement$. Since $\symHashValueMax$ is decreasing with the number of processed elements, the stop condition is satisfied very quickly for later elements, often already after the first generated point as for $\symInputElement_4$ in this example. The resulting signature is $\symSignature=(\symInputElement_3, \symInputElement_1, \symInputElement_1, \symInputElement_2)$.}
\label{fig:example}
\end{figure*}

We also present a performance optimization for ProbMinHash1 and ProbMinHash3 leading to corresponding equivalent algorithms ProbMinHash1a and ProbMinHash3a, respectively.  The interleaved processing of input elements by using an additional buffer can significantly improve the performance for medium input sizes $\symInputSize$.

We investigated specializations of all our algorithms on the conventional Jaccard similarity $\symJaccard$, where we only have binary weights $\symWeight(\symInputElement)\in\{ 0, 1\}$. In this case, ProbMinHash1, ProbMinHash1a, and ProbMinHash2 are statistically equivalent to the original minwise hashing approach and can be used, in contrast to the state of the art, as drop-in replacements. A very interesting algorithm is also the specialization of ProbMinHash3a, which results in a very fast algorithm for $\symJaccard$ with a time complexity of $\symBigO(\symInputSize + \symHashSize\log\symHashSize)$ and a space complexity of $\symBigO(\symHashSize\log\symHashSize)$. 

We conducted rigorous experiments using synthetic data to investigate the runtime behavior as well as the estimation error for different input and signature sizes. The results are in accordance with the theoretical considerations. The source code required to reproduce the results presented in this paper has been published on GitHub at \url{https://github.com/oertl/probminhash}.

\section{Methodology}

The computation of signatures for $\symJaccardP$ as defined by \eqref{equ:signature_def} and \eqref{equ:def_hash_jp} involves $\symInputSize\symHashSize$ hash values. $\symHashSize$ hash values, one for each signature component, must be calculated per element. The elements with the smallest hash values finally define the signature. Therefore, while processing elements, the minimum hash values seen so far must be kept for each signature component. In \cref{alg:p_min_hash} the array $(\symHashMin_1,\symHashMin_2,\ldots,\symHashMin_\symHashSize)$ is used for that purpose. The maximum of all those minimum hash values $\symHashValueMax:=\max_\symHashIndex  \symHashMin_\symHashIndex$ defines a limit, which is decreasing with the number of processed elements, and beyond which hash values of any further elements will not have an impact on the final signature. Thus, if we were able to track this limit efficiently over time, and at the same time, if we were able to generate hash values of an element in ascending order, processing of that element could be stopped as soon as its hash values exceed the current value of $\symHashValueMax$.

To produce the hash values $\symHashValue_\symHashIndex(\symInputElement)$ of a given element $\symInputElement$ in ascending order we propose to pick them from a positive monotonic increasing random sequence $\symSequencePoint(\symInputElement)=(\symPoint_1(\symInputElement),\symPoint_2(\symInputElement),\ldots)$ of points with random labels from $\{1,2,\ldots,\symHashSize\}$. Let $\symLabel_\symIndex(\symInputElement)$ denote the label of point $\symPoint_\symIndex(\symInputElement)$ and $\symSequenceIndices(\symInputElement)=(\symLabel_1(\symInputElement),\symLabel_2(\symInputElement),\ldots)$ be the corresponding sequence of labels. We define $\symHashValue_\symHashIndex(\symInputElement)$ as the first point of sequence $\symSequencePoint(\symInputElement)$ with a label equal to $\symHashIndex$:
\begin{equation}
\label{equ:def_new_exp}
\symHashValue_\symHashIndex(\symInputElement):=\symPoint_{\min\{\symIndex:\symLabel_\symIndex(\symInputElement) = \symHashIndex\}}(\symInputElement)
=
\min_{\symIndex:\symLabel_\symIndex(\symInputElement) = \symHashIndex}\symPoint_\symIndex(\symInputElement).
\end{equation}
The sequences $\symSequencePoint(\symInputElement)$ and $\symSequenceIndices(\symInputElement)$ are generated using a \ac{PRNG} that is initialized with $\symInputElement$ as seed to have independence between different elements. If the sequences are chosen such that the hash values $\symHashValue_\symHashIndex(\symInputElement)$ satisfy \eqref{equ:def_hash_jp} for all $\symHashIndex\in\{1,2,\ldots,\symHashSize\}$, they can be used in \eqref{equ:signature_def} to obtain signatures for $\symJaccardP$. As shown later, this requires, among other things, that the point density in $\symSequencePoint(\symInputElement)$ depends on the weight $\symWeight(\symInputElement)$.
  
\myAlg{
\caption{Basic structure of ProbMinHash algorithms.}
\label{alg:prob_min_hash}
\KwIn{$\symWeight$}
\KwOut{$\symSignature_1,
\symSignature_2,
\ldots,
\symSignature_{\symHashSize}$}
$(
\symHashMin_1,
\symHashMin_2,
\ldots,
\symHashMin_{\symHashSize}
)\gets
(\infty,\infty,\ldots,\infty)$\;
\ForAll{$\symInputElement\in\symUniverse$ such that $\symWeight(\symInputElement) > 0$}
{
$\symWeightInverse\gets 1/\symWeight(\symInputElement)$\;
$\symRNG \gets$ \New \acs{PRNG} with seed $\symInputElement$\;
$\symHashValue\gets\text{generate first element of $\symSequencePoint(\symInputElement)$ using $\symRNG$}$ \;
$\symIndex\gets1$\;
\While{$\symHashValue < \symHashValueMax$}{
    $\symHashIndex\gets\text{generate $\symIndex$-th element of $\symSequenceIndices(\symInputElement)$ using $\symRNG$}$\;
    \If{$\symHashValue < \symHashMin_\symHashIndex$}
    {
        $\symHashMin_\symHashIndex \gets \symHashValue$\;
        $\symSignature_\symHashIndex \gets \symInputElement$\;
        $\symHashValueMax\gets\max_\symHashIndex  \symHashMin_\symHashIndex$\;
        \lIf{$\symHashValue \geq \symHashValueMax$}{\Break}
    }
    $\symIndex\gets\symIndex+1$\;
    $\symHashValue\gets\text{generate $\symIndex$-th element of $\symSequencePoint(\symInputElement)$ using $\symRNG$}$ \;
}
}
}

\cref{alg:prob_min_hash} shows the signature calculation using this technique. Elements of sequences $\symSequencePoint(\symInputElement)$ and $\symSequenceIndices(\symInputElement)$ are lazily generated using the \ac{PRNG} $\symRNG$ which is initialized using $\symInputElement$ as seed. As soon as points $\symPoint_\symIndex(\symInputElement)$ are greater than or equal to $\symHashValueMax$ processing of element $\symInputElement$ can be stopped. For the very first element $\symInputElement$, when $\symHashValueMax$ is still infinite, this stop condition is satisfied as soon as all possible labels have appeared at least once in $\symSequenceIndices(\symInputElement)$. For further elements, $\symHashValueMax$ has already become smaller and the stop condition will likely be satisfied much earlier. If $\symInputSize\gg \symHashSize$, $\symHashValueMax$ will get very small such that the while-loop can be entirely skipped for most elements and a huge speedup factor in the order of $\symHashSize$ can be expected compared to \cref{alg:p_min_hash}. \cref{fig:example} demonstrates \cref{alg:prob_min_hash} for an example with $\symHashSize=4$ and a set of 4 weighted elements.

\subsection{Sequence Generation}
\label{sec:sequencegeneration}

If the hash values $\symHashValue_\symHashIndex(\symInputElement)$ satisfy \eqref{equ:def_hash_jp}, they must be identically distributed. This implies that labels are uniformly distributed and therefore $\symProbability(\symLabel_\symIndex(\symInputElement) = \symHashIndex) = \frac{1}{\symHashSize}$ for all $\symHashIndex\in\{ 1,2,\ldots,\symHashSize\}$. This can be achieved by sampling the labels $\symLabel_\symIndex(\symInputElement)$ either from the multiset $\symMultiSet_\symMultiplicity:=\{1^\symMultiplicity 2^\symMultiplicity \ldots  \symHashSize^\symMultiplicity\}$ without replacement or from $\{1,2,\ldots, \symHashSize\}$ with replacement. In the following, sampling with replacement is treated as a special case of sampling without replacement where $\symMultiplicity\rightarrow\infty$.
Given $\symMultiplicity$, which defines the multiset $\symMultiSet_\symMultiplicity$ used for generating $\symSequenceIndices(\symInputElement)$, we need to find an appropriate monotonic increasing random sequence $\symSequencePoint(\symInputElement)$ such that \eqref{equ:def_new_exp} satisfies \eqref{equ:def_hash_jp}. We propose two different methods which we refer to as uncorrelated and correlated generation of $\symSequencePoint(\symInputElement)$, respectively.

The uncorrelated approach uses a well known property about exponential spacings described in \cite{Devroye1986} which allows sampling of $\symHashSize$ independent random values from an exponential distribution with rate parameter equal to 1 in ascending order by using the recursion $\symPoint_{\symIndex}(\symInputElement)\sim\symPoint_{\symIndex-1}(\symInputElement) + \frac{1}{\symHashSize - \symIndex+1}\symExponential(1)$ and $\symPoint_0(\symInputElement):=0$. Hence, the $\symIndex$-th largest random value is simply obtained by adding a value drawn from an exponential distribution with rate $(\symHashSize - \symIndex+1)$ to the $(\symIndex-1)$-th largest random value. 
We apply this to generate the sequence $\symSequencePoint(\symInputElement)$ by sampling $\symHashSize\symMultiplicity$ independent and exponentially distributed random values with rate parameter $\symRate\symWeight(\symInputElement)/\symMultiplicity$ in ascending order. The corresponding recursion formula is obtained by replacing $\symHashSize$ by $\symHashSize\symMultiplicity$ and scaling with factor $\symMultiplicity/(\symRate\symWeight(\symInputElement))$
\begin{equation}
    \label{equ:uncorrelated_recursion}
    \symPoint_{\symIndex}(\symInputElement)\sim \symPoint_{\symIndex-1}(\symInputElement) + \frac{1}{\symRate\symWeight(\symInputElement)}\frac{\symMultiplicity}{\symHashSize\symMultiplicity - \symIndex+1}\symExponential(1).
\end{equation}
Since the label sequence $\symSequenceIndices(\symInputElement)$ is produced by sampling without replacement from $\symMultiSet_\symMultiplicity$, there will always be exactly $\symMultiplicity$ points with the same label. The minimum of those $\symMultiplicity$ points, which are exponentially distributed with rate $\symRate\symWeight(\symInputElement)/\symMultiplicity$, is exponentially distributed with rate $\symRate\symWeight(\symInputElement)$, because the minimum of independent exponentially distributed random values is also exponentially distributed and the rate is equal to the sum of rates of all contributing random values \cite{Ross2019}. As a consequence, $\symHashValue_\symHashIndex(\symInputElement)$ as defined in \eqref{equ:def_new_exp} is exponentially distributed with rate $\symRate\symWeight(\symInputElement)$ and satisfies \eqref{equ:def_hash_jp} as required. By construction, the hash values of an element and therefore also the resulting signature components \eqref{equ:signature_def} are mutually independent and the variance of estimator \eqref{equ:jaccard_estimator} will be the same as for P-MinHash given by $\symJaccardP(1-\symJaccardP)/\symHashSize$.

For the correlated approach, we consider the probability $\symFirstOccurP_\symIndex:=\symProbability(\min(\{\symIndexJ:\symLabel_\symIndexJ(\symInputElement)=\symHashIndex\})=\symIndex)$ that $\symHashValue_\symHashIndex(\symInputElement)$ as defined by \eqref{equ:def_new_exp} is given by the $\symIndex$-th point $\symPoint_\symIndex(\symInputElement)$. This corresponds to the probability that the $\symIndex$-th label $\symLabel_\symIndex(\symInputElement)$ is the first label with value $\symHashIndex$.
Since the labels are drawn without replacement from $\symMultiSet_\symMultiplicity$, the probability that $\symLabel_\symIndexJ(\symInputElement)=\symHashIndex$ is $\frac{\symMultiplicity}{\symHashSize\symMultiplicity-\symIndexJ+1}$ given that all $(\symIndexJ-1)$ previous labels are different from $\symHashIndex$. 
Therefore, the probability that the first $(\symIndex-1)$ labels are different from $\symHashIndex$ and the $\symIndex$-th label is equal to $\symHashIndex$ is given by
\begin{equation}
    \label{equ:correlated_probabilities}
    \symFirstOccurP_\symIndex
    =
    \prod_{\symIndexJ=1}^{\symIndex-1}
    \left(1 - \frac{\symMultiplicity}{\symHashSize\symMultiplicity-\symIndexJ+1}
    \right)
    \cdot
    \frac{\symMultiplicity}{\symHashSize\symMultiplicity-\symIndex+1}.
\end{equation}
$\symFirstOccurP_\symIndex$ is zero for $\symIndex > \symHashSize\symMultiplicity-\symMultiplicity+1$, because latest after sampling $\symHashSize\symMultiplicity-\symMultiplicity+1$ labels without replacement from $\symMultiSet_\symMultiplicity$, each label has been sampled at least once. By nature, we have $\sum_{\symIndex=1}^{\symHashSize\symMultiplicity-\symMultiplicity+1} \symFirstOccurP_\symIndex = 1$.
$\symHashValue_\symHashIndex(\symInputElement)$ is given by the $\symIndex$-th point $\symPoint_\symIndex(\symInputElement)$ with probability $\symFirstOccurP_\symIndex$. Therefore, if $\symPoint_\symIndex(\symInputElement)$ is sampled from the two-sided truncated exponential distribution
\begin{align}
\symPoint_\symIndex(\symInputElement)
&\sim
\symExponential(\symRate\symWeight(\symInputElement); \symBoundary_{\symIndex-1}, \symBoundary_\symIndex)
\nonumber
\\
&\sim\symBoundary_{\symIndex-1} + \symExponential(\symRate\symWeight(\symInputElement); 0, \symBoundary_\symIndex - \symBoundary_{\symIndex-1})\label{equ:correlated_points}
\\
&\sim\symBoundary_{\symIndex-1} + (\symBoundary_\symIndex - \symBoundary_{\symIndex-1})\symExponential(\symRate\symWeight(\symInputElement) (\symBoundary_\symIndex - \symBoundary_{\symIndex-1}); 0, 1)
\nonumber
\end{align}
with rate $\symRate\symWeight(\symInputElement)$ and support $[\symBoundary_{\symIndex-1}, \symBoundary_\symIndex)$,
where the boundaries $\symBoundary_\symIndex$ are chosen such that $\symProbability(\symBoundary_{\symIndex-1} \leq \symPoint < \symBoundary_{\symIndex})  = \symFirstOccurP_\symIndex$ with $\symBoundary_0:=0$, $\symHashValue_\symHashIndex(\symInputElement)$ will be exponentially distributed with rate $\symRate\symWeight(\symInputElement)$ by construction. 
Since the points $\symPoint_\symIndex(\symInputElement)$ are sampled from disjoint intervals, $\symHashValue_\symHashIndex(\symInputElement)$ will not be independent for different $\symHashIndex$. As a consequence, the variance of estimator \eqref{equ:jaccard_estimator} will differ from \eqref{equ:jaccard_variance}, in contrast to uncorrelated generation of $\symSequencePoint(\symInputElement)$. If $\symInputSize$ is in the range of $\symHashSize$ or smaller, the frequency of elements in the signature is usually not very balanced. Correlated generation results in more evenly distributed points, because there will always be exactly one single point in each interval $[\symBoundary_{\symIndex-1},\symBoundary_{\symIndex})$. This leads to a more balanced frequency distribution of elements in the signature in case the input size $\symInputSize$ is in the order of $\symHashSize$ or smaller. The similarity estimates using \eqref{equ:jaccard_estimator} will be more accurate and the variance will be lower than for P-MinHash or the correlated approach, respectively, as confirmed by our experimental results presented later.

Although the choice of $\symMultiplicity$ is arbitrary, we consider only two cases to be practical. The first is $\symMultiplicity=1$, because it minimizes the number of points that need to be generated. The second case is $\symMultiplicity\rightarrow\infty$. Even though this means that there is no worst case upper bound for the number of required points, it offers the advantage of sampling with replacement for $\symSequenceIndices(\symInputElement)$. Sampling with replacement is less expensive, because it does not require an algorithm like Fisher-Yates shuffling which must incorporate the sampling history \cite{Fisher1938}. The two cases $\symMultiplicity=1$ and $\symMultiplicity\rightarrow\infty$ combined either with uncorrelated or correlated generation of $\symSequencePoint(\symInputElement)$ result in four different algorithms which will be described and analyzed in more detail in \crefrange{sec:probminhash1}{sec:probminhash4}.
But first we have to explain the last missing part of our approach, namely how $\symHashValueMax$ is efficiently tracked in \cref{alg:prob_min_hash}. 

\subsection{Stop Limit Update}
\myAlg{
\caption{
Maintenance of stop limit
$\symHashValueMax:=\max(
\symHashMin_1,
\symHashMin_2,
\ldots,
\symHashMin_{\symHashSize})$. $\symHashMin_{\symHashSize+1},\ldots,\symHashMin_{2\symHashSize-1}$ are the parent nodes of a binary tree spanned over $\symHashMin_1,
\symHashMin_2,
\ldots,
\symHashMin_{\symHashSize}$. $\symHashMin_{\symHashSize + \lceil \symHashIndex/2 \rceil}$ is the parent of $\symHashMin_\symHashIndex$. If a leaf node $\symHashMin_\symHashIndex$ with $\symHashIndex \in \{1,2,\ldots,\symHashSize\}$ is replaced by a smaller value $\symHashValue < \symHashMin_\symHashIndex$, following procedure updates the root node $\symHashMin_{2\symHashSize-1}$ which corresponds to $\symHashValueMax$.
}
\label{alg:track_max}
\KwIn{$\symHashValue$, $\symHashIndex$} 
\While{$\symHashValue < \symHashMin_\symHashIndex$}{
    $\symHashMin_\symHashIndex\gets\symHashValue$\;
    $\symIndex\gets\symHashSize + \lceil \symHashIndex / 2 \rceil$\Com*{index of parent}
    \lIf{$\symIndex\geq 2\symHashSize$}{\Break}
    $\symIndexJ\gets ((\symHashIndex-1)\mathbin{\symXor} 1) + 1$\Com*{index of sibling, bitwise $\symXor$ operation}
    \lIf{$\symHashMin_\symIndexJ\geq \symHashMin_\symIndex$}{\Break}
    \lIf{$\symHashValue<\symHashMin_\symIndexJ$}{$\symHashValue\gets\symHashMin_\symIndexJ$}
    $\symHashIndex\gets\symIndex$\;
}
}
The naive calculation of $\symHashValueMax$ by iterating over the array $(\symHashMin_1,\symHashMin_2,\ldots,\symHashMin_{\symHashSize})$ every time one of its values has changed takes $\symBigO(\symHashSize)$ time and would make our approach very slow. 
Therefore we need a data structure that is able to update $\symHashValueMax$ in sublinear time if the value of some $\symHashMin_\symHashIndex$ is replaced by a smaller value. A max-heap data structure cannot be used, because it only gives access to the maximum and does not allow fast updates of other elements. The solution is a binary tree constructed over the array $(\symHashMin_1,\symHashMin_2,\ldots,\symHashMin_{\symHashSize})$. The parent nodes are stored in the same array $\symHashMin$ by expanding it to the size $2\symHashSize-1$. The value of a parent node is defined to be the maximum of the values of both child nodes. By definition, the value of the root node will be $\symHashValueMax$. If some value $\symHashMin_\symHashIndex$ is replaced by some smaller hash value $\symHashValue$, \cref{alg:track_max} can be used to update the tree including $\symHashValueMax$. Starting from the modified leaf node, the algorithm makes a bottom-up tree traversal until no further change is necessary. It is a slightly modified but equivalent version of the algorithm presented in \cite{Ertl2018}. 

Clearly, the worst case time complexity is $\symBigO(\log \symHashSize)$. However, for ProbMinHash where the values of labels $\symLabel_\symIndex(\symInputElement)$ are equally frequent, the expected time complexity is $\symBigO(1)$, which can be explained as follows: Assume $\symLabel_\symIndex(\symInputElement) = \symHashIndex$ and hence $\symHashMin_\symHashIndex$ has been chosen to be potentially updated by $\symPoint_\symIndex(\symInputElement)$. Obviously, the probability that the value of $\symHashMin_\symHashIndex$ is actually changed is at most 1. The probability that the parent of $\symHashMin_\symHashIndex$ is modified, is at most $\frac{1}{2}$, because it is equally likely that the parent value, which is the maximum of the values of its two children, is given by the sibling of $\symHashMin_\symHashIndex$. Decrementing the value of $\symHashMin_\symHashIndex$ has no impact on the parent in this case. Continuation of this argumentation shows that the expected number of node updates must be bounded by the geometric series $1 +  \frac{1}{2} + \frac{1}{4} + \ldots = 2$ and is therefore constant. 

\subsection{ProbMinHash1}
\label{sec:probminhash1}

Sampling with replacement ($\symMultiplicity\rightarrow\infty$) for $\symSequenceIndices(\symInputElement)$ together with uncorrelated sampling for $\symSequencePoint(\symInputElement)$ are the ingredients for the ProbMinHash1 algorithm. $\symMultiplicity\rightarrow\infty$ and the choice $\symRate:=\frac{1}{\symHashSize}$ for the free parameter $\symRate$ simplifies the recursion \eqref{equ:uncorrelated_recursion} to $\symPoint_{\symIndex}(\symInputElement) \sim\symPoint_{\symIndex-1}(\symInputElement) + \frac{1}{\symWeight(\symInputElement)}\symExponential(1)$. The specialization of \cref{alg:prob_min_hash} for this case is shown as \cref{alg:prob_min_hash_1}.

To analyze the runtime behavior we consider the number of points that need to be generated for each element $\symInputElement_\symIndexJ$ to satisfy the stop condition $\symHashValue\geq \symHashValueMax$. For the first element $\symInputElement_1$, the stop condition is fulfilled as soon as all possible labels have appeared in $\symSequenceIndices(\symInputElement_1)$. As this corresponds to the coupon collector's problem \cite{Cormen2009, Mitzenmacher2017}, this takes $\symHashSize \symHarmonic_ \symHashSize$ points on average, where $\symHarmonic_\symHashSize:=1+\frac{1}{2}+\ldots+\frac{1}{\symHashSize}$ denotes the $\symHashSize$-th harmonic number. 
Processing of the second element $\symInputElement_2$ already takes less time, because $\symHashValueMax$ has decreased and is no longer infinite. Consider the point sequence obtained by combining and sorting $\symSequencePoint(\symInputElement_1)$ and $\symSequencePoint(\symInputElement_2)$ together with the corresponding joint label sequence given by $\symSequenceIndices(\symInputElement_1)$ and $\symSequenceIndices(\symInputElement_2)$. 
Again, the stop condition is satisfied as soon as all possible labels have shown up in the combined label sequence. As before, the expected sequence length is $\symHashSize\symHarmonic_\symHashSize$. However, since 
the density of points that come from $\symSequencePoint(\symInputElement_1)$ and $\symSequencePoint(\symInputElement_2)$ is proportional to $\symWeight(\symInputElement_1)$ and $\symWeight(\symInputElement_2)$, respectively, we expect that the proportion of points originating from $\symInputElement_2$ is $\symWeight(\symInputElement_2)/(\symWeight(\symInputElement_1) + \symWeight(\symInputElement_2))$. Therefore, the expected number of generated points for $\symInputElement_2$ is $\symHashSize\symHarmonic_\symHashSize \symWeight(\symInputElement_2) / (\symWeight(\symInputElement_1) + \symWeight(\symInputElement_2))$. Continuing in this way leads to $\symHashSize \symHarmonic_\symHashSize  \symWeight(\symInputElement_\symIndexJ) /\sum_{\symIndex=1}^\symIndexJ \symWeight(\symInputElement_\symIndex)$ for $\symInputElement_\symIndexJ$. 
If we assume that elements are not processed in any particular order with respect to their weights, the expectation of $\symWeight(\symInputElement_\symIndexJ)/
\sum_{\symIndex=1}^\symIndexJ \symWeight(\symInputElement_\symIndex)$ is equal to $\frac{1}{\symIndexJ}$. 
Summation over all $\symInputSize$ input elements yields the expected total number of generated points which is $\symHashSize\symHarmonic_\symHashSize \symHarmonic_\symInputSize$. Using $\symHarmonic_\symHashSize = \symBigO(\log\symHashSize)$ and incorporating the fixed costs $\symBigO(\symInputSize)$ associated with each of all $\symInputSize$ processed elements, the amortized overall time complexity is $\symBigO(\symInputSize + \symHashSize (\log \symHashSize)(\log \symInputSize)) \leq \symBigO(\symInputSize + \symHashSize \log^2 \symHashSize)$. The inequality means that there exists a constant $\symBigConstant$ such that $\symInputSize + \symHashSize (\log \symHashSize)(\log \symInputSize) \leq \symBigConstant (\symInputSize + \symHashSize \log^2 \symHashSize)$ holds for all $\symHashSize\geq 1$ and $\symInputSize\geq 1$. In particular, we have been able to prove this inequality for $\symBigConstant = \frac{3}{2}.$

\myAlg{
\caption{ProbMinHash1.}
\label{alg:prob_min_hash_1}
\KwIn{$\symWeight$}
\KwOut{$\symSignature_1,
\symSignature_2,
\ldots,
\symSignature_{\symHashSize}$}
$(
\symHashMin_1,
\symHashMin_2,
\ldots,
\symHashMin_{\symHashSize}
)\gets
(\infty,\infty,\ldots,\infty)$\;
\ForAll{$\symInputElement\in\symUniverse$ such that $\symWeight(\symInputElement) > 0$}
{
$\symWeightInverse\gets 1/\symWeight(\symInputElement)$\;
$\symRNG \gets$ \New \acs{PRNG} with seed $\symInputElement$\;
$\symHashValue\gets \symWeightInverse\cdot\symRNG[\symExponential(1)]$\;
\While{$\symHashValue < \symHashValueMax$}{
    $\symHashIndex\gets\symRNG[\symUniform(\{1,2,\ldots,\symHashSize\})]$\;
    \If{$\symHashValue < \symHashMin_\symHashIndex$}
    {
        $\symHashMin_\symHashIndex \gets \symHashValue$\;
        $\symSignature_\symHashIndex \gets \symInputElement$\;
        update $\symHashValueMax$ using \cref{alg:track_max}\;
        \lIf{$\symHashValue \geq \symHashValueMax$}{\Break}
    }
    $\symHashValue\gets \symHashValue + \symWeightInverse\cdot\symRNG[\symExponential(1)]$\;
}
}
}

Theoretically, if all elements are processed in ascending order with respect to their weights and the sequence $\symWeight(\symInputElement_\symIndexJ)/\sum_{\symIndex=1}^{\symIndexJ} \symWeight(\symInputElement_\symIndex)$ is rather constant than decreasing like $\frac{1}{\symIndexJ}$, the time complexity would become $\symBigO(\symInputSize\symHashSize (\log \symHashSize) )$ in the worst case. 
However, if the data is expected to be ordered, it is unlikely that the data will have to be processed in streaming mode, since most likely a sorting step has been performed beforehand. Therefore, to avoid the worst-case behavior, the elements could either be processed in reverse order so that the weights decrease, or, if the data fits in memory, the elements could be processed in random order using Fisher-Yates shuffling \cite{Fisher1938} which is an $\symBigO(\symInputSize)$-operation.

\subsection{ProbMinHash2}

Our second algorithm uses sampling without replacement ($\symMultiplicity=1$) for $\symSequenceIndices(\symInputElement)$ and uncorrelated sampling for $\symSequencePoint(\symInputElement)$. Unfortunately, sampling without replacement is more expensive than sampling with replacement. It is usually done using Fisher-Yates shuffling which requires an array $\symPermutationArray=(\symPermutationArray_1,\symPermutationArray_2,\ldots,\symPermutationArray_\symHashSize)$ of size $\symHashSize$ with initial values $\symPermutationArray_\symIndex=\symIndex$ \cite{Fisher1938}. Due to the stop condition, ProbMinHash only needs the labels of a few points for most input elements. Since the $\symBigO(\symHashSize)$ allocation and initialization costs would lead to an $\symBigO(\symInputSize\symHashSize)$ algorithm, we propose \cref{alg:permutation}, a variant of Fisher-Yates shuffling. It reuses the array $\symPermutationArray$ to amortize allocation costs. Furthermore, it applies lazy initialization by using a permutation counter $\symVersionCounter$ and an additional array $\symVersionArray$ that indicates already initialized array components of $\symPermutationArray$ for the current permutation. $\symVersionArray_\symIndex = \symVersionCounter$ means that $\symPermutationArray_\symIndex$ has already been initialized. Otherwise, $\symPermutationArray_\symIndex$ is considered to be equal to its initial value $\symIndex$. To start a new permutation for the next input element it is sufficient to increment the counter which is just an $\symBigO(1)$ operation.

\myAlg{
\caption{Lazy generation of random permutation elements based on Fisher-Yates shuffling.}
\label{alg:permutation}
\Function{\InitPermutationGenerator{$\symHashSize$}}{
allocate array $(\symPermutationArray_1,\symPermutationArray_2,\ldots,\symPermutationArray_\symHashSize)$\;
$(\symVersionArray_1,\symVersionArray_2,\ldots,\symVersionArray_\symHashSize)\gets(0,0,\ldots,0)$\;
$\symVersionCounter\gets0$\;
}
\Function{\ResetPermutationGenerator{}}{
$\symIndex\gets0$\;
$\symVersionCounter\gets \symVersionCounter + 1$\;
}
\Function{\GenerateNextPermutationElement{$\symRNG$}}{
$\symIndex\gets \symIndex + 1$\;
$\symIndexJ\gets\symIndex + \symRNG[\symUniform(\{0, 1,\ldots,\symHashSize-\symIndex\})]$\;
$\symHashIndex\gets\begin{cases}\symPermutationArray_\symIndexJ & \symVersionArray_\symIndexJ = \symVersionCounter\\ \symIndexJ & \symVersionArray_\symIndexJ\neq \symVersionCounter \end{cases}$\;
$\symPermutationArray_\symIndexJ\gets\begin{cases}\symPermutationArray_\symIndex & \symVersionArray_\symIndex = \symVersionCounter\\ \symIndex & \symVersionArray_\symIndex\neq \symVersionCounter \end{cases}$\;
$\symVersionArray_\symIndexJ\gets \symVersionCounter$\;
\KwRet $\symHashIndex$\;
}
}

\myAlg{
\caption{ProbMinHash2.}
\label{alg:prob_min_hash_2}
\KwIn{$\symWeight$}
\KwOut{$\symSignature_1,
\symSignature_2,
\ldots,
\symSignature_{\symHashSize}$}
$(
\symHashMin_1,
\symHashMin_2,
\ldots,
\symHashMin_{\symHashSize}
)\gets
(\infty,\infty,\ldots,\infty)$\;
\InitPermutationGenerator{\symHashSize}\;
\ForAll{$\symInputElement\in\symUniverse$ such that $\symWeight(\symInputElement) > 0$}
{
$\symWeightInverse\gets 1/\symWeight(\symInputElement)$\;
$\symRNG \gets$ \New \acs{PRNG} with seed $\symInputElement$\;
\ResetPermutationGenerator{}\;
$\symIndex\gets 1$\;
$\symHashValue\gets \symWeightInverse\cdot\symRNG[\symExponential(1)]$\;
\While{$\symHashValue < \symHashValueMax$}{
    $\symHashIndex\gets\GenerateNextPermutationElement{\symRNG}$\;
    \If(\Com*[f]{always satisfied, if $\symIndex = \symHashSize$}){$\symHashValue < \symHashMin_\symHashIndex$}
    {   
        $\symHashMin_\symHashIndex \gets \symHashValue$\;
        $\symSignature_\symHashIndex \gets \symInputElement$\;
        update $\symHashValueMax$ using \cref{alg:track_max}\;
        \lIf(\Com*[f]{always satisfied, if $\symIndex = \symHashSize$}){$\symHashValue \geq \symHashValueMax$}{\Break}
    }
    $\symIndex\gets\symIndex+1$\;
    $\symHashValue\gets \symHashValue + \symWeightInverse\cdot\symBeta_\symIndex\cdot \symRNG[\symExponential(1)]$\Com*{$\symBeta_\symIndex:=\symHashSize/(\symHashSize-\symIndex+1)$}
}
}
}

The free parameter in \eqref{equ:uncorrelated_recursion} is chosen again as $\symRate:=\frac{1}{\symHashSize}$. In this way the recursion simplifies together with $\symMultiplicity=1$  to $\symPoint_{\symIndex}(\symInputElement)\sim\symPoint_{\symIndex-1}(\symInputElement) + \frac{1}{\symWeight(\symInputElement)}\frac{\symHashSize}{\symHashSize - \symIndex+1}\symExponential(1)$ and the very first and most frequently calculated point is simply given by $\symPoint_1(\symInputElement)\sim\frac{1}{\symWeight(\symInputElement)}\symExponential(1)$. The factors $\symBeta_\symIndex:=\frac{\symHashSize}{\symHashSize - \symIndex+1}$ can be precomputed to avoid the costly floating-point divisions. ProbMinHash2 as a whole is shown as \cref{alg:prob_min_hash_2}. 
Sampling the labels from $\symMultiSet_1=\{1,2,\ldots,\symHashSize\}$ without replacement guarantees that the stop condition will always be satisfied after generating $\symHashSize$ points. The last point will always be greater than or equal to $\symHashValueMax$. Therefore, the worst case time complexity is bounded by $\symBigO(\symInputSize\symHashSize)$ which equals the time complexity of P-MinHash. In contrast, there is no such worst case upper bound for ProbMinHash1.  It is obvious that the expected time complexity of ProbMinHash2 is bounded by that of ProbMinHash1, because ProbMinHash2 generates at most one point per label and thus fewer points, which according to \eqref{equ:def_new_exp} have no influence on $\symHashValue_\symHashIndex(\symInputElement)$.

\subsection{ProbMinHash3}

The third algorithm combines sampling with replacement $(\symMultiplicity\rightarrow\infty)$ for label generation and the correlated sampling approach for $\symSequencePoint(\symInputElement)$. According to \eqref{equ:correlated_probabilities}
$\symFirstOccurP_\symIndex
=
\frac{1}{\symHashSize}(1-\frac{1}{\symHashSize})^{\symIndex-1}$ as $\symMultiplicity\rightarrow\infty$. The interval boundaries $\symBoundary_{\symIndex}$ that satisfy $\symProbability(\symBoundary_{\symIndex-1} \leq \symPoint < \symBoundary_{\symIndex})  = \symFirstOccurP_\symIndex$, if $\symPoint$ is exponentially distributed with rate $\symRate\symWeight(\symInputElement)$, are given by $\symBoundary_\symIndex = \frac{\symIndex}{\symRate\symWeight(\symInputElement)}\log(1+\frac{1}{\symHashSize-1})$. By choosing $\symRate:=\log(1+\frac{1}{\symHashSize-1})$, which requires $\symHashSize \geq 2$, we have $\symBoundary_\symIndex =\frac{\symIndex}{\symWeight(\symInputElement)}$ and the points can be generated according to \eqref{equ:correlated_points} using $\symPoint_\symIndex(\symInputElement) \sim\frac{\symIndex-1}{\symWeight(\symInputElement)} + \frac{1}{\symWeight(\symInputElement)}\symExponential(\symRate;0,1)$. 
ProbMinHash3 is shown as \cref{alg:prob_min_hash_3}. 
In contrast to ProbMinHash1 and ProbMinHash2, the stop condition within the inner loop makes use of the fact that the $\symIndex$-th point $\symPoint_\symIndex(\symInputElement)$ is sampled from $[\symBoundary_{\symIndex-1},\symBoundary_{\symIndex})=[\frac{\symIndex-1}{\symWeight(\symInputElement)},\frac{\symIndex}{\symWeight(\symInputElement)})$. Therefore, if the lower bound of this interval is already equal to or greater than $\symHashValueMax$, the stop condition will be satisfied in any case. The generation of $\symPoint_\symIndex(\symInputElement)$ can therefore be omitted.
The time complexity is similar to that of ProbMinHash1. The same argumentation can be used to derive the upper bounds $\symBigO(\symInputSize + \symHashSize (\log\symHashSize)(\log\symInputSize)) \leq \symBigO(\symInputSize + \symHashSize \log^2 \symHashSize)$.

\subsection{ProbMinHash4}
\label{sec:probminhash4}

\myAlg[!t]{
\caption{ProbMinHash3, requires $\symHashSize\geq2$.}
\label{alg:prob_min_hash_3}
\KwIn{$\symWeight$}
\KwOut{$\symSignature_1,
\symSignature_2,
\ldots,
\symSignature_{\symHashSize}$}
$(
\symHashMin_1,
\symHashMin_2,
\ldots,
\symHashMin_{\symHashSize}
)\gets
(\infty,\infty,\ldots,\infty)$\;
\ForAll{$\symInputElement\in\symUniverse$ such that $\symWeight(\symInputElement) > 0$}
{
$\symWeightInverse\gets 1/\symWeight(\symInputElement)$\;
$\symRNG \gets$ \New \acs{PRNG} with seed $\symInputElement$\;
$\symHashValue\gets \symWeightInverse\cdot\symRNG[\symExponential(\symRate;0,1)]$\Com*{$\symRate:=\log(1+1/(\symHashSize-1))$}
$\symIndex\gets 1$\;
\While{$\symHashValue < \symHashValueMax$}{
    $\symHashIndex\gets\symRNG[\symUniform(\{1,2,\ldots,\symHashSize\})]$\;
    \If{$\symHashValue < \symHashMin_\symHashIndex$}
    {
        $\symHashMin_\symHashIndex \gets \symHashValue$\;
        $\symSignature_\symHashIndex \gets \symInputElement$\;
        update $\symHashValueMax$ using \cref{alg:track_max}\;
    }
    $\symIndex\gets\symIndex+1$\;
    $\symHashValue\gets\symWeightInverse\cdot(\symIndex-1)$\;
    \lIf{$\symHashValue \geq \symHashValueMax$}{\Break}
    $\symHashValue\gets \symHashValue + \symWeightInverse\cdot\symRNG[\symExponential(\symRate;0,1)]$\;
}
}
}

\myAlg{
\caption{ProbMinHash4, requires $\symHashSize\geq2$.}
\label{alg:prob_min_hash_4}
\KwIn{$\symWeight$}
\KwOut{$\symSignature_1,
\symSignature_2,
\ldots,
\symSignature_{\symHashSize}$}
$(
\symHashMin_1,
\symHashMin_2,
\ldots,
\symHashMin_{\symHashSize}
)\gets
(\infty,\infty,\ldots,\infty)$\;
\InitPermutationGenerator{\symHashSize}\;
\ForAll{$\symInputElement\in\symUniverse$ such that $\symWeight(\symInputElement) > 0$}
{
$\symWeightInverse\gets 1/\symWeight(\symInputElement)$\;
$\symRNG \gets$ \New \acs{PRNG} with seed $\symInputElement$\;
\ResetPermutationGenerator{}\;
$\symHashValue\gets \symWeightInverse\cdot\symRNG[\symExponential(\symRate_1;0,1)]$\Com*{$\symRate_\symIndex:=\log(1+1/(\symHashSize-\symIndex))$}
$\symIndex\gets 1$\;
\While{$\symHashValue < \symHashValueMax$}{
    $\symHashIndex\gets\GenerateNextPermutationElement{\symRNG}$\;
    \If{$\symHashValue < \symHashMin_\symHashIndex$}
    {   
        $\symHashMin_\symHashIndex \gets \symHashValue$\;
        $\symSignature_\symHashIndex \gets \symInputElement$\;
        update $\symHashValueMax$ using \cref{alg:track_max}\;
    }
    $\symIndex\gets\symIndex+1$\;
    \lIf(\Com*[f]{$\symGamma_{\symIndex}:=\log(1+\symIndex/(\symHashSize-\symIndex))/\symRate_1$}){$\symWeightInverse\cdot\symGamma_{\symIndex-1} \geq \symHashValueMax$}{\Break}
    \uIf{$\symIndex < \symHashSize$}{
    $\symHashValue\gets \symWeightInverse\cdot\left(\symGamma_{\symIndex-1} + (\symGamma_{\symIndex}-\symGamma_{\symIndex-1})\cdot\symRNG[\symExponential(\symRate_{\symIndex};0,1)]\right)$\;
    }
    \Else{
        $\symHashValue\gets \symWeightInverse\cdot(\symGamma_{\symHashSize-1} + \symDelta \cdot \symRNG[\symExponential(1)])$\Com*{$\symDelta:=1/\symRate_1$}
        \If(\Com*[f]{$\symHashValue < \symHashValueMax$ and $\symIndex = \symHashSize$ imply $\symHashValue < \symHashMin_\symHashIndex$}){$\symHashValue < \symHashValueMax$}
        {   
            $\symHashIndex\gets\GenerateNextPermutationElement{\symRNG}$\;
            $\symHashMin_\symHashIndex \gets \symHashValue$\;
            $\symSignature_\symHashIndex \gets \symInputElement$\;
            update $\symHashValueMax$ using \cref{alg:track_max}\;
        }
        \Break\;
    }
}
}
}

The fourth variant combines sampling without replacement $(\symMultiplicity=1)$ for label generation and correlated sampling for $\symSequencePoint(\symInputElement)$. In this case \eqref{equ:correlated_probabilities} reduces to $\symFirstOccurP_\symIndex=\frac{1}{\symHashSize}$ and the interval boundaries are given by $\symBoundary_\symIndex =\frac{1}{\symRate\symWeight(\symInputElement)}\log(1+\frac{\symIndex}{\symHashSize-\symIndex})$. 
Setting again $\symRate:=\log(1+\frac{1}{\symHashSize-1})$ 
simplifies \eqref{equ:correlated_points} for the first and most frequently computed point to $\symPoint_1(\symInputElement)\sim \frac{1}{\symWeight(\symInputElement)}\symExponential(\symRate;0,1)$.
More generally, we obtain $\symPoint_\symIndex(\symInputElement)  \sim\frac{1}{\symWeight(\symInputElement)}
(\symGamma_{\symIndex-1} + (\symGamma_{\symIndex} - \symGamma_{\symIndex-1})\symExponential(\symRate_\symIndex;0,1))$ with $\symRate_\symIndex:=\log(1+1/(\symHashSize-\symIndex))$ and $\symGamma_{\symIndex}:=\log(1+\symIndex/(\symHashSize-\symIndex))/\symRate_1$ for $\symIndex < \symHashSize$ and $\symPoint_\symHashSize(\symInputElement)  \sim\frac{1}{\symWeight(\symInputElement)}(\symGamma_{\symHashSize-1} + \symDelta\symExponential(1))$ with $\symDelta:=1/\symRate_1$. 
ProbMinHash4 is shown as \cref{alg:prob_min_hash_4}.
Similar to ProbMinHash2, \cref{alg:permutation} is used for sampling without replacement and the number of generated points is limited by $\symHashSize$ resulting in a  worst case time complexity of $\symBigO(\symInputSize\symHashSize)$. Like for ProbMinHash3, the first stop condition within the while-loop makes use of the fact that the $\symIndex$-th point $\symPoint_\symIndex(\symInputElement)$ originates from $[\symBoundary_{\symIndex-1},\symBoundary_{\symIndex})$. Since ProbMinHash4 avoids the generation of points with the same labels, its expected time complexity is bounded by that of ProbMinHash3.

\subsection{Interleaved Element Processing}
\label{sec:interleaved}

The presented algorithms process elements sequentially. Processing the first elements is much more time-consuming than later elements when $\symHashValueMax$ has already decreased and the stop condition is satisfied earlier. In the best case, if $\symHashValueMax$ is already small enough, processing of elements can be terminated immediately when checking the stop condition the first time.
In any case, the first point $\symPoint_1(\symInputElement)$ of each element $\symInputElement$ must be generated. Therefore it makes sense to do what needs to be done first and to calculate only the first points of all elements in a first pass. Elements, that do not satisfy the stop condition right after the first generated point, need to be stored in a buffer, because further points may contribute to the signature. In the next pass we take the elements from the buffer, generate and process the second smallest points $\symPoint_2(\symInputElement)$, and, if the stop condition is still not fulfilled, reinsert the elements into the buffer. This procedure is repeated until the buffer is finally empty.

The performance gain will be most significant for intermediate input sizes $\symInputSize$, because we do not expect any speedup for both extreme cases, $\symInputSize=1$ and $\symInputSize\rightarrow\infty$. When processing just a single input element, interleaving is not possible and the same number of random points must be generated. For $\symInputSize\rightarrow\infty$, $\symHashValueMax$ will be small enough for most except the first elements anyway, so that the stop condition is almost always satisfied by their first points. The relative performance gain is negligible in this case.

The space requirements are given by the maximum number of elements that are simultaneously stored in the buffer. Obviously, this maximum is given by the number of elements in the buffer right after the first pass. The actual memory costs are composed by the elements and the corresponding states required for further point generation. This includes \ac{PRNG} states, and in case of sampling without replacement, the states of the shuffling algorithm. It is not feasible to store an array of length $\symHashSize$ for each element in the buffer as needed by shuffling algorithms. Therefore, a more complex data structure like a hash table that only stores initialized elements of that array is needed. We did a couple of experiments using hash tables, but found that the additional hash table lookups outweigh the performance gain due to interleaved processing. Therefore, we focused on ProbMinHash1 and ProbMinHash3 which are both based on sampling with replacement and which do not rely on the sampling history. ProbMinHash1a and ProbMinHash3a are the corresponding logically equivalent algorithms using interleaved processing and are shown as \cref{alg:prob_min_hash_1a} and \cref{alg:prob_min_hash_3a}, respectively.

To analyze the space complexity we assume that elements $\symInputElement_\symIndexJ$ are not processed in a sorted order with regard to their weights. As a consequence, the first points $\symPoint_1(\symInputElement_\symIndexJ)$ of different elements $\symInputElement_\symIndexJ$ can be considered to be identically distributed.
The expected maximum buffer size is given by the sum of individual probabilities that an element is added to the buffer during the first pass.

For ProbMinHash1a an element $\symInputElement_\symIndexJ$ is added to the buffer, if its first point $\symPoint_1(\symInputElement_\symIndexJ)$ is smaller than $\symHashValueMax$. Since it takes $\symHashSize\symHarmonic_\symHashSize$ elements on average, according to the coupon collector's problem \cite{Cormen2009, Mitzenmacher2017}, until the labels of their first points $\symLabel_1(\symInputElement_\symIndexJ)$ cover all possible label values $\{ 1,2,\ldots,\symHashSize\}$, $\symHashValueMax$ is roughly given by the $\symHashSize\symHarmonic_\symHashSize$-th smallest point seen so far. The probability, that the first point of the $\symIndexJ$-th element $\symPoint_1(\symInputElement_\symIndexJ)$ is among the $\symHashSize\symHarmonic_\symHashSize$ smallest points, is given by 
$\min(1, \symHashSize\symHarmonic_\symHashSize/\symIndexJ)$. As this corresponds to the probability that $\symPoint_1(\symInputElement_\symIndexJ)$ is smaller than $\symHashValueMax$ and that $\symInputElement_\symIndexJ$ is inserted into the buffer, summation of these probabilities for all $\symInputSize$ elements finally yields the expected buffer size $\sum_{\symIndexJ=1}^\symInputSize \min(1, \symHashSize\symHarmonic_\symHashSize/\symIndexJ) = \symBigO(\symHashSize (\log\symHashSize)(\log\symInputSize))$. The expected time complexity of ProbMinHash1a is $\symBigO(\symInputSize + \symHashSize (\log\symHashSize)(\log\symInputSize))$, because the space complexity and the number of processed elements $\symInputSize$ are obvious lower bounds while the time complexity of ProbMinHash1 is an obvious upper bound.

\myAlg{
\caption{ProbMinHash1a.}
\label{alg:prob_min_hash_1a}
\KwIn{$\symWeight$}
\KwOut{$\symSignature_1,
\symSignature_2,
\ldots,
\symSignature_{\symHashSize}$}
$(
\symHashMin_1,
\symHashMin_2,
\ldots,
\symHashMin_{\symHashSize}
)\gets
(\infty,\infty,\ldots,\infty)$\;
$\symBuffer\gets\text{empty dynamic array}$\Com*{initialize buffer}
$\symBufferCounter \gets 0$\Com*{number of elements in buffer}
\ForAll{$\symInputElement\in\symUniverse$ such that $\symWeight(\symInputElement) > 0$}
{
$\symWeightInverse\gets 1/\symWeight(\symInputElement)$\;
$\symRNG \gets$ \New \acs{PRNG} with seed $\symInputElement$\;
$\symHashValue\gets \symWeightInverse\cdot\symRNG[\symExponential(1)]$\;
\lIf{$\symHashValue \geq \symHashValueMax$}{\Continue}
$\symHashIndex\gets\symRNG[\symUniform(\{1,2,\ldots,\symHashSize\})]$\;
\If{$\symHashValue < \symHashMin_\symHashIndex$}
{
    $\symHashMin_\symHashIndex \gets \symHashValue$\;
    $\symSignature_\symHashIndex \gets \symInputElement$\;
    update $\symHashValueMax$ using \cref{alg:track_max}\;
    \lIf{$\symHashValue \geq \symHashValueMax$}{\Continue}
}
$\symBufferCounter\gets\symBufferCounter+1$\;
$\symBuffer_\symBufferCounter\gets(\symInputElement, \symWeightInverse, \symRNG, \symHashValue)$\;
}
\While{$\symBufferCounter>0$}{
    $\symBufferCounterTwo\gets0$\;
    \For{$\symIndexJ\gets 1,2,\ldots,\symBufferCounter$}{
        $(\symInputElement, \symWeightInverse, \symRNG, \symHashValue)\gets\symBuffer_\symIndexJ$\;
        \lIf{$\symHashValue \geq \symHashValueMax$}{\Continue}
        $\symHashValue\gets \symHashValue + \symWeightInverse\cdot\symRNG[\symExponential(1)]$\;
        \lIf{$\symHashValue \geq \symHashValueMax$}{\Continue}
        $\symHashIndex\gets\symRNG[\symUniform(\{1,2,\ldots,\symHashSize\})]$\;
        \If{$\symHashValue < \symHashMin_\symHashIndex$}
        {
            $\symHashMin_\symHashIndex \gets \symHashValue$\;
            $\symSignature_\symHashIndex \gets \symInputElement$\;
            update $\symHashValueMax$ using \cref{alg:track_max}\;
            \lIf{$\symHashValue \geq \symHashValueMax$}{\Continue}
        }
        $\symBufferCounterTwo\gets\symBufferCounterTwo+1$\;
        $\symBuffer_{\symBufferCounterTwo}\gets(\symInputElement, \symWeightInverse, \symRNG, \symHashValue)$\;
    
    }
    $\symBufferCounter\gets\symBufferCounterTwo$\;
}

}

\myAlg[t]{
\caption{ProbMinHash3a, requires $\symHashSize\geq2$.}
\label{alg:prob_min_hash_3a}
\KwIn{$\symWeight$}
\KwOut{$\symSignature_1,
\symSignature_2,
\ldots,
\symSignature_{\symHashSize}$}
$(
\symHashMin_1,
\symHashMin_2,
\ldots,
\symHashMin_{\symHashSize}
)\gets
(\infty,\infty,\ldots,\infty)$\;
$\symBuffer\gets\text{empty dynamic array}$\Com*{initialize buffer}
$\symBufferCounter \gets 0$\Com*{number of elements in buffer}
\ForAll{$\symInputElement\in\symUniverse$ such that $\symWeight(\symInputElement) > 0$}
{
$\symWeightInverse\gets 1/\symWeight(\symInputElement)$\;
$\symRNG \gets$ \New \acs{PRNG} with seed $\symInputElement$\;
$\symHashValue\gets \symWeightInverse\cdot \symRNG[\symExponential(\symRate;0,1)]$\Com*{$\symRate:=\log(1+1/(\symHashSize-1))$}
\lIf{$\symHashValue \geq \symHashValueMax$}{\Continue}
$\symHashIndex\gets\symRNG[\symUniform(\{1,2,\ldots,\symHashSize\})]$\;
\If{$\symHashValue < \symHashMin_\symHashIndex$}
{
    $\symHashMin_\symHashIndex \gets \symHashValue$\;
    $\symSignature_\symHashIndex \gets \symInputElement$\;
    update $\symHashValueMax$ using \cref{alg:track_max}\;
}
\lIf{$\symWeightInverse \geq \symHashValueMax$}{\Continue}
$\symBufferCounter\gets\symBufferCounter+1$\;
$\symBuffer_\symBufferCounter\gets(\symInputElement, \symWeightInverse, \symRNG)$\;
}
$\symIndex\gets 2$\;
\While{$\symBufferCounter>0$}{
    $\symBufferCounterTwo\gets0$\;
    \For{$\symIndexJ\gets 1,2,\dots,\symBufferCounter$}{
        $(\symInputElement, \symWeightInverse, \symRNG)\gets\symBuffer_\symIndexJ$\;
        $\symHashValue\gets \symWeightInverse\cdot(\symIndex-1)$\;
        \lIf{$\symHashValue \geq \symHashValueMax$}{\Continue}
        
        $\symHashValue\gets \symHashValue + \symWeightInverse\cdot\symRNG[\symExponential(\symRate;0,1)]$\;
        \lIf{$\symHashValue \geq \symHashValueMax$}{\Continue}
        $\symHashIndex\gets\symRNG[\symUniform(\{1,2,\ldots,\symHashSize\})]$\;
        \If{$\symHashValue < \symHashMin_\symHashIndex$}
        {
            $\symHashMin_\symHashIndex \gets \symHashValue$\;
            $\symSignature_\symHashIndex \gets \symInputElement$\;
            update $\symHashValueMax$ using \cref{alg:track_max}\;
        }
        \lIf{$\symWeightInverse\cdot \symIndex\geq \symHashValueMax$}{\Continue}
        $\symBufferCounterTwo\gets\symBufferCounterTwo+1$\;
        $\symBuffer_{\symBufferCounterTwo}\gets(\symInputElement, \symWeightInverse, \symRNG)$\;
    
    }
    $\symBufferCounter\gets\symBufferCounterTwo$\;
    $\symIndex\gets\symIndex+1$\;
}

}

For ProbMinHash3a we first consider the unweighted case with all weights equal to 1. If we are in the $\symIndex$-th pass, when all points are sampled from $[\symIndex-1,\symIndex)$, the stop condition is always satisfied and elements are not (re)inserted into the buffer as soon as $\symHashValueMax\leq\symIndex$. On average, this will be the case after generating the first $\symHashSize\symHarmonic_\symHashSize=\symBigO(\symHashSize\log \symHashSize)$ points, because $\symHashValueMax$ is approximately given by the $\symHashSize\symHarmonic_\symHashSize$-th smallest point as before. The processing time of the remaining elements, which then immediately fulfill the stop condition, is naturally limited by $\symBigO(\symInputSize)$, resulting in an overall time complexity of $\symBigO(\symInputSize + \symHashSize\log \symHashSize)$.
The expected maximum number of elements in the buffer after the first pass is given by $\min(\symInputSize, \symHashSize\symHarmonic_\symHashSize)\leq\symBigO(\symHashSize\log \symHashSize)$. It is remarkable that the space complexity does not depend on the input size $\symInputSize$ in contrast to ProbMinHash1a. 

In the general case, the space complexity of ProbMinHash3a depends on the distribution of weights $\symDistribution_\symWeight(\symY)= \symProbability(\symWeight\leq \symY)$. Based on our experimental results and the following theoretical considerations, we assume that there is an upper bound that is independent of $\symInputSize$ as long as $\symDistribution_\symWeight(\symY)$ has a power-tail with index $\symTailIndex> 1$ which means that  $1-\symDistribution_\symWeight(\symY)\sim \symY^{-\symTailIndex}$ as $\symY\rightarrow 0$ and that the mean is finite. The first points $\symPoint_1$ of elements given the weight $\symWeight$ are distributed as $\symPoint_1\vert\symWeight\sim\frac{1}{\symWeight}\symExponential(\symRate_1;0,1)$ as mentioned in \cref{sec:probminhash4}. It can be shown that the power-tail of  $\symDistribution_\symWeight$ yields an asymptotic lower bound for the distribution function $\symDistribution_{\symPoint_1}(\symY)=\symProbability(\symPoint_1 \leq \symY)\geq \symY\symExpectation(\symWeight)\symBigConstant$ as $\symY\rightarrow 0$ where $\symExpectation(\symWeight)$ denotes the average weight and $\symBigConstant$ is some constant. $\symHashValueMax$ is again roughly given by the $\symHashSize\symHarmonic_\symHashSize$-smallest point. Therefore, after inserting the first $\symIndexJ$ elements, $\symHashValueMax$ is approximated by the $(\symHashSize\symHarmonic_\symHashSize/\symIndexJ)$-quantile $\symHashValueMax\approx\symDistribution^{-1}_{\symPoint_1}(\symHashSize\symHarmonic_\symHashSize/\symIndexJ)\leq \frac{\symHashSize\symHarmonic_\symHashSize}{\symIndexJ \symBigConstant \symExpectation(\symWeight)}$ as $\symIndexJ\rightarrow\infty$. An element is added to the buffer only if $\symWeightInverse<\symHashValueMax$. The corresponding probability is $\symProbability(1/\symWeight<\symHashValueMax) = 1 - \symDistribution_\symWeight(1/\symHashValueMax)\leq 1- \symDistribution_\symWeight(\symIndexJ \symBigConstant \symExpectation(\symWeight)/( \symHashSize\symHarmonic_\symHashSize))\propto \symIndexJ^{-\symTailIndex}$ as $\symIndexJ\rightarrow\infty$. Since the hyperharmonic series $\sum_{\symIndexJ=1}^\infty \symIndexJ^{-\symTailIndex}$ converges for $\symTailIndex>1$, the sum of probabilities, that elements are added to the buffer, has an upper bound independent of $\symInputSize$. 

\subsection{Truncated Exponential Sampling}

ProbMinHash3 and ProbMinHash4 require the generation of many random values that follow a truncated exponential distribution $\symExponential(\symRate;0,1)$. The straightforward approach, inverse transform sampling, requires the evaluation of a logarithm. We have not found any other method in literature that avoids, like the ziggurat method for the exponential distribution \cite{Marsaglia2000}, expensive function calls. Therefore, we propose the following method based on rejection sampling.

\begin{figure}[t]
\centering
\includegraphics[width=\columnwidth]{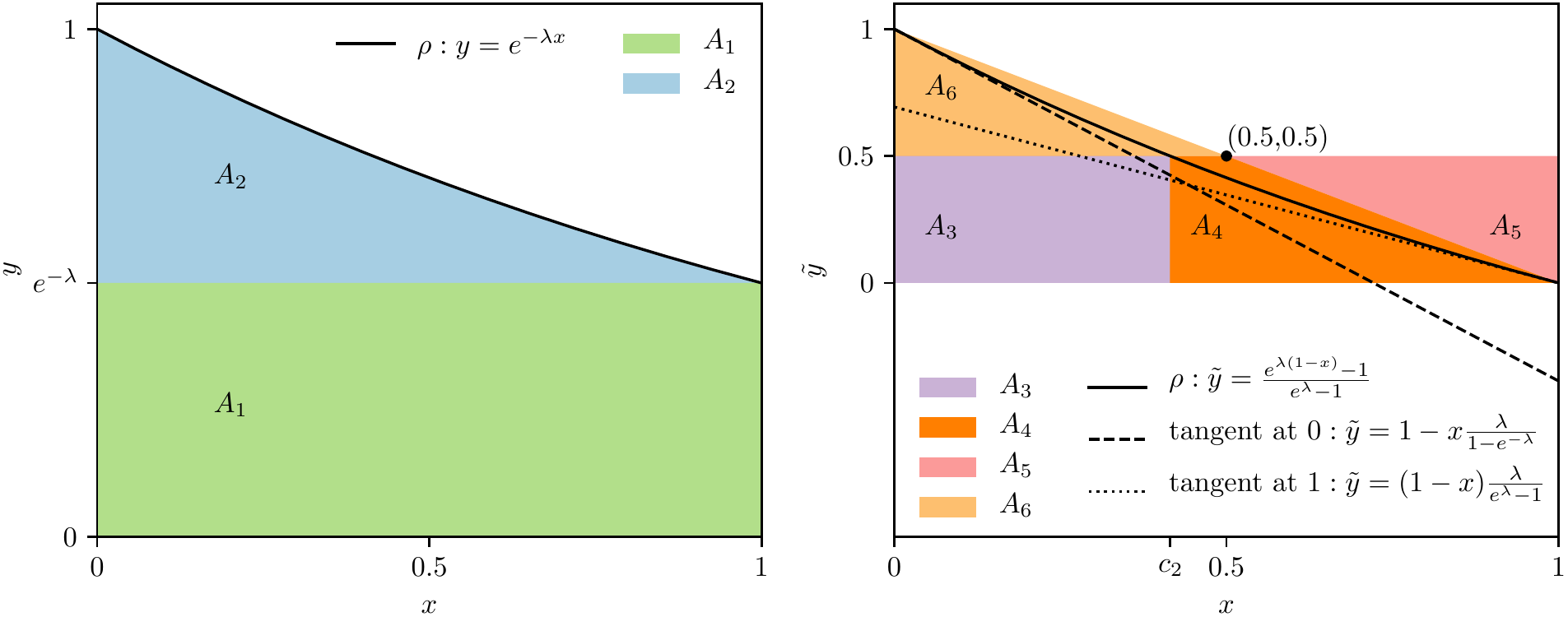}
\caption{Sampling from a truncated exponential distribution $\symExponential(\symRate;0,1)$.}
\label{fig:exp_truncated}
\end{figure}

\cref{fig:exp_truncated} shows the function $\symDensity(\symPoint)= e^{-\symRate\symPoint}$ which is proportional to the probability density function of $\symExponential(\symRate;0,1)$. Clearly, if we sample points $(\symPoint,\symY)$ uniformly from the region below $\symDensity$ with $\symPoint\in[0,1)$ and $\symY\leq\rho(\symPoint)$, $\symPoint$ will be distributed as $\symExponential(\symRate;0,1)$. The region is split into $\symArea_1 = [0,1)\times[0,e^{-\symRate})$ and the remaining part $\symArea_2$. 
In order to achieve uniformity a point is sampled from $\symArea_1$ with propability $|\symArea_1|/|\symArea_1\cup\symArea_2|$. Otherwise, a point is taken from $\symArea_2$. The area sizes are given by $|\symArea_1| =e^{-\symRate}$ and $|\symArea_1\cup\symArea_2|=(1-e^{-\symRate})/\symRate$. To realize the corresponding Bernoulli trial we generate a uniformly distributed random value $\symPoint\in[0, \symConstant_1)$ with $\symConstant_1:=|\symArea_1\cup\symArea_2|/|\symArea_1| = (e^{\symRate}-1)/\symRate$. Since $\symProbability(\symPoint<1) = |\symArea_1|/|\symArea_1\cup\symArea_2|$, we sample a point from $\symArea_1$ if $\symPoint<1$. In this case $\symPoint$ is uniformly distributed over $[0,1)$, and $\symPoint$ can therefore be reused as the $\symPoint$-coordinate of the point we want to generate. Because $\symArea_1$ is an axis-aligned rectangle, all $\symPoint$-coordinates are equally likely, so generating the $\symY$-coordinate can be omitted, and $\symPoint$ can be returned directly as result.

Rejection sampling is used to get a point from $\symArea_2$ in the other case, when $\symPoint\geq 1$. For that we first introduce a different $\symY$-scale defined by the transformation $\tilde{\symY} = (\symY - e^{-\symRate})/(1-e^{-\symRate})$ as shown in the right part of \cref{fig:exp_truncated}. Instead from $\symArea_2$, we sample points from the triangle $\symArea_3\cup\symArea_4\cup\symArea_6$ and reject those which are not below $\symDensity$. To do so we sample from the rectangle $\symArea_3\cup\symArea_4\cup\symArea_5 = [0,1)\times[0,0.5)$ which has the same area as the triangle, because the areas of $\symArea_5$ and $\symArea_6$ are equal. In case a sampled point belongs to $\symArea_5$, it is mapped to $\symArea_6$ by reflection at point $(0.5, 0.5)$. Next we need to test whether the point is below $\symDensity$. In order to avoid the expensive exponential function evaluation, we first check whether it is either below the tangent at 0 or 1 given by $\tilde{\symY}=1-\symPoint\frac{\symRate}{1-e^{-\symRate}}$ and $\tilde{\symY}=(1-\symPoint)\frac{\symRate}{e^{\symRate}-1}$, respectively. Since $\symDensity$ is convex, the point can be accepted in both cases. A test against $\symDensity$ is only needed in the remaining case. If the point is finally accepted, its $\symPoint$-coordinate is returned as result. 
An additional performance optimization can be  introduced when sampling from the rectangle $\symArea_3\cup\symArea_4\cup\symArea_5$. We first sample its $\symPoint$-coordinate from $[0,1)$. If it is less than $\symConstant_2:=\log(2/(1+e^{-\symRate}))/\symRate$, the point belongs to  $\symArea_3$ which is the part of the rectangle entirely below $\symDensity$. Therefore, a point in $\symArea_3$ can be immediately accepted without considering its $\symY$-coordinate.

\cref{alg:truncated_exp} summarizes the whole procedure. It is especially efficient for small $\symRate$, because then the first if-condition $\symPoint<1$ is satisfied with high probability which allows an early termination. For ProbMinHash3 we need random values from $\symExponential(\symRate;0,1)$ with rate  $\symRate = \log(1+\frac{1}{\symHashSize-1})$. Therefore $\symProbability(\symPoint<1)=\symRate/(e^\symRate-1)=(\symHashSize-1)\log(1+\frac{1}{\symHashSize-1})\sim 1-\frac{1}{2(\symHashSize-1)}$ as $\symHashSize\rightarrow \infty$. Since $\symHashSize$ is typically in the hundreds or even greater, drawing a value from $\symExponential(\symRate;0,1)$ is almost as cheap as generating a uniform random value. For ProbMinHash4, the rate parameter increases with the number of points from $\symRate_1=\log(1+\frac{1}{\symHashSize-1})$, which is the same as for ProbMinHash3, to $\symRate_{\symHashSize-1} = \log(2)$ for the second last point $\symPoint_{\symHashSize-1}(\symInputElement)$. Even in this worst case the first if-condition is still satisfied with a probability of $\symProbability(\symPoint<1)=\log(2) \approx 69.3\%$.

\myAlg[t]{
\caption{Fast generation of random values from $\symExponential(\symRate; 0, 1)$.}
\label{alg:truncated_exp}
\KwIn{$\symRate$, $\symRNG$}
\KwOut{$\symPoint$}
$\symPoint\gets \symConstant_1\cdot\symRNG[\symUniform([0,1))]$\Com*{$\symConstant_1:=(e^{\symRate}-1)/\symRate$}
\lIf(\Com*[f]{sample from $\symArea_1$}){$\symPoint < 1$}{\Return $\symPoint$}
\Loop(\Com*[f]{otherwise, start rejection sampling from $\symArea_2$}){
    $\symPoint\gets \symRNG[\symUniform([0,1))]$\Com*{sample $\symPoint$ from $\symArea_3\cup\symArea_4\cup\symArea_5$}
    \lIf(\Com*[f]{point is in $\symArea_3$, $\symConstant_2:=\log(2/(1+e^{-\symRate}))/\symRate$}){$\symPoint < \symConstant_2$}{\Return $\symPoint$}
    $\tilde{\symY}\gets 0.5\cdot\symRNG[\symUniform([0,1))]$\Com*{sample $\tilde{\symY}$ from $\symArea_3\cup\symArea_4\cup\symArea_5$}
    \If(\Com*[f]{point is in $\symArea_5$}){$\tilde{\symY} > 1- \symPoint$}{
        $(\symPoint,\tilde{\symY})\gets (1-\symPoint, 1-\tilde{\symY})$\Com*{map from $\symArea_5$ to $\symArea_6$ by reflection at $(0.5,0.5)$}
    }
    \lIf(\Com*[f]{below tangent at 0, $\symConstant_3:=(1-e^{-\symRate})/\symRate$}){$\symPoint\leq \symConstant_3\cdot(1-\tilde{\symY})$}{\Return $\symPoint$}
    \lIf(\Com*[f]{below tangent at 1}){$\tilde{\symY}\cdot\symConstant_1 \leq 1- \symPoint$}{\Return $\symPoint$}
    \lIf(\Com*[f]{below $\symDensity$}){$\tilde{\symY}\cdot\symConstant_1 \cdot \symRate \leq \exp(\symRate\cdot (1-\symPoint))-1$}{\Return $\symPoint$}
}
}

\subsection{Conventional Jaccard Similarity}
\label{sec:conventional_jaccard}

Since the probability Jaccard similarity $\symJaccardP$ is a generalization of the conventional Jaccard similarity $\symJaccard$, all ProbMinHash algorithms can also be used to calculate signatures for $\symJaccard$ by allowing only binary weights $\symWeight(\symInputElement)\in\{0,1\}$. In this section we investigate whether the corresponding specializations lead to novel or already known locality-sensitive hash algorithms for the conventional Jaccard similarity $\symJaccard$.

For ProbMinHash1 and ProbMinHash2 the restriction to binary weights means that there is no need to compute $\symWeightInverse$ because it is always 1. As a consequence, floating-point multiplications with $\symWeightInverse$ can be saved. In addition, ProbMinHash1a benefits from a smaller memory footprint, because $\symWeightInverse$ does not need to be stored in the buffer. Since ProbMinHash1, ProbMinHash1a, and ProbMinHash2 are all based on uncorrelated point generation as discussed in \cref{sec:sequencegeneration}, they are statistically equivalent to the original MinHash algorithm in contrast to other state-of-the-art algorithms like \ac{OPH} \cite{Li2012}, SuperMinHash \cite{Ertl2017}, or \ac{FSS} \cite{Dahlgaard2017}. Therefore they can be used as faster drop-in replacements without changing any statistical properties. This is especially important, if signatures are used in the context of locality-sensitive hashing \cite{Bawa2005, Indyk1998, Lv2007} where signature components are usually considered to be independent.

ProbMinHash3 and ProbMinHash4, which use correlated point generation, benefit even more from the restriction to binary weights. The interval boundaries $\symBoundary_\symIndex$ are the same for all elements in this case. This means that the $\symIndex$-th largest points of different elements are identically distributed. Since only the relative order of points plays a role according to \eqref{equ:def_new_exp}, we can use any other continuous distribution. For the sake of simplicity and also for performance reasons we take the uniform distribution. This means, we replace $\symRNG[\symExponential(\symRate;0,1)]$ by $\symRNG[\symUniform([0,1))]$ in ProbMinHash3 and ProbMinHash3a. In ProbMinHash4 we set $\symGamma_\symIndex=\symIndex$ and substitute both $\symRNG[\symExponential(\symRate_{\symIndex};0,1)]$ and $\symDelta \cdot\symRNG[\symExponential(1)]$ by $\symRNG[\symUniform([0,1))]$.

Although the ProbMinHash algorithms are new for the general case, some specialized variants for the conventional Jaccard similarity $\symJaccard$ relate to already known algorithms. ProbMinHash1 corresponds to the BagMinHash1 algorithm \cite{Ertl2018} if both are fully specialized for the unweighted case. ProbMinHash1a has some similarities to BagMinHash2 which uses a min-heap for buffering elements that are still able to contribute to the signature while processing them in ascending order of their last points. In contrast, ProbMinHash1a calculates the $\symIndex$-th smallest points of all relevant elements in the $\symIndex$-th pass regardless of the preceding points.

In the unweighted case, the for-loop of ProbMinHash3a in \cref{alg:prob_min_hash_3a} is similar to \ac{OPH} without densification \cite{Li2012} and to the first iteration of \ac{FSS} \cite{Dahlgaard2017}, respectively. Thus, ProbMinHash3a is nearly equivalent, if its while-loop does not contribute to the signature, which is typically the case for large input sizes $\symInputSize \gg \symHashSize$. However, ProbMinHash3a additionally tracks $\symHashValueMax$, allowing earlier termination. In this way the generation of the first label $\symLabel_1(\symInputElement)$ and also the access to the corresponding signature component can often be avoided, which results in a small advantage in performance. In contrast, the other algorithms must always randomly select and access at least one signature component for each element. The while-loop in \cref{alg:prob_min_hash_3a} can be regarded as a new alternative densification scheme for \ac{OPH} \cite{Mai2019,Shrivastava2017,Shrivastava2014a}. At the expense of a buffer of size $\symBigO(\symHashSize \log\symHashSize)$ the estimation error for $\symJaccard$ is significantly reduced for small input sizes, as our experimental results will show. 

ProbMinHash4 for the unweighted case corresponds to the SuperMinHash algorithm \cite{Ertl2017}. 
The main difference is again the stop condition. SuperMinHash only tracks a histogram of points which only allows discrete stop limits. ProbMinHash4 keeps track of $\symHashValueMax$, which is slightly more expensive. However, it pays off for larger input sizes, because the stop limit can go below 1. As a consequence, the label generation can be avoided in many cases leading again to a slightly better performance. As ProbMinHash4 is logically equivalent to the SuperMinHash algorithm it also shares the theoretically proven better variance of the Jaccard similarity estimator given by \eqref{equ:superminhash_variance}. 
           
\section{Experiments}
All algorithms were implemented using C++. The corresponding source code and the Python scripts to reproduce the results and figures shown in the following are available on Github at \url{https://github.com/oertl/probminhash}. The \ac{PRNG}, which is used by the presented algorithms, is particularly important for good performance, since it is called in the innermost loops. For our theoretical considerations we have assumed an ideal random number generator. Therefore, the output of the chosen \ac{PRNG} should be indistinguishable from that. Poor randomness would lead to estimation errors significantly different from the theoretical predictions. 
We used the Wyrand algorithm (version 4) which was recently developed and published on GitHub \cite{Yi2019}. It is very fast, has a state of 64 bits, and passes a series of statistical quality tests. Seeded with a 64-bit hash value of the input element, it produces a sequence of 64-bit pseudo-random integers. We consume random bits very economically. For example, to generate a double-precision floating-point number from $[0,1)$ 53 random bits corresponding to the significand precision are sufficient. Only if all 64 bits are consumed, the next bunch of 64 bits will be generated. 

We would like to point out that the use of a weak \ac{PRNG} could potentially cause our algorithms to run endlessly. On the one hand, random values are partly generated using rejection sampling techniques for which the number of trials follows a geometric distribution and on the other hand, ProbMinHash1 and ProbMinHash3 which are based on sampling with replacement require that all possible label values appear at least once to satisfy the stop condition for the first element. 
We have never experienced an endless iteration with our implementation so far. However, if this is a concern, the number of iterations of the corresponding loops should be explicitly limited. The limits should be chosen in a way that, assuming an ideal random number generator, they are only reached with a negligible probability. For example, the probability that any label will not appear among the first $(\symMu\,\symHashSize \log\symHashSize)$ points is less than $\symHashSize^{1-\symMu}$ \cite{Mitzenmacher2017}.

All algorithms including P-MinHash were implemented with the same methods of random value generation to allow a fair comparison. The ziggurat method \cite{Marsaglia2000} as implemented in the Boost C++ libraries \cite{Boost} was used to generate exponentially distributed random values. We applied \cref{alg:truncated_exp} for truncated exponential distributions. Uniform random integers were produced as described in \cite{Lemire2019}.

\begin{figure*}[t]
\centering
\includegraphics[width=\textwidth]{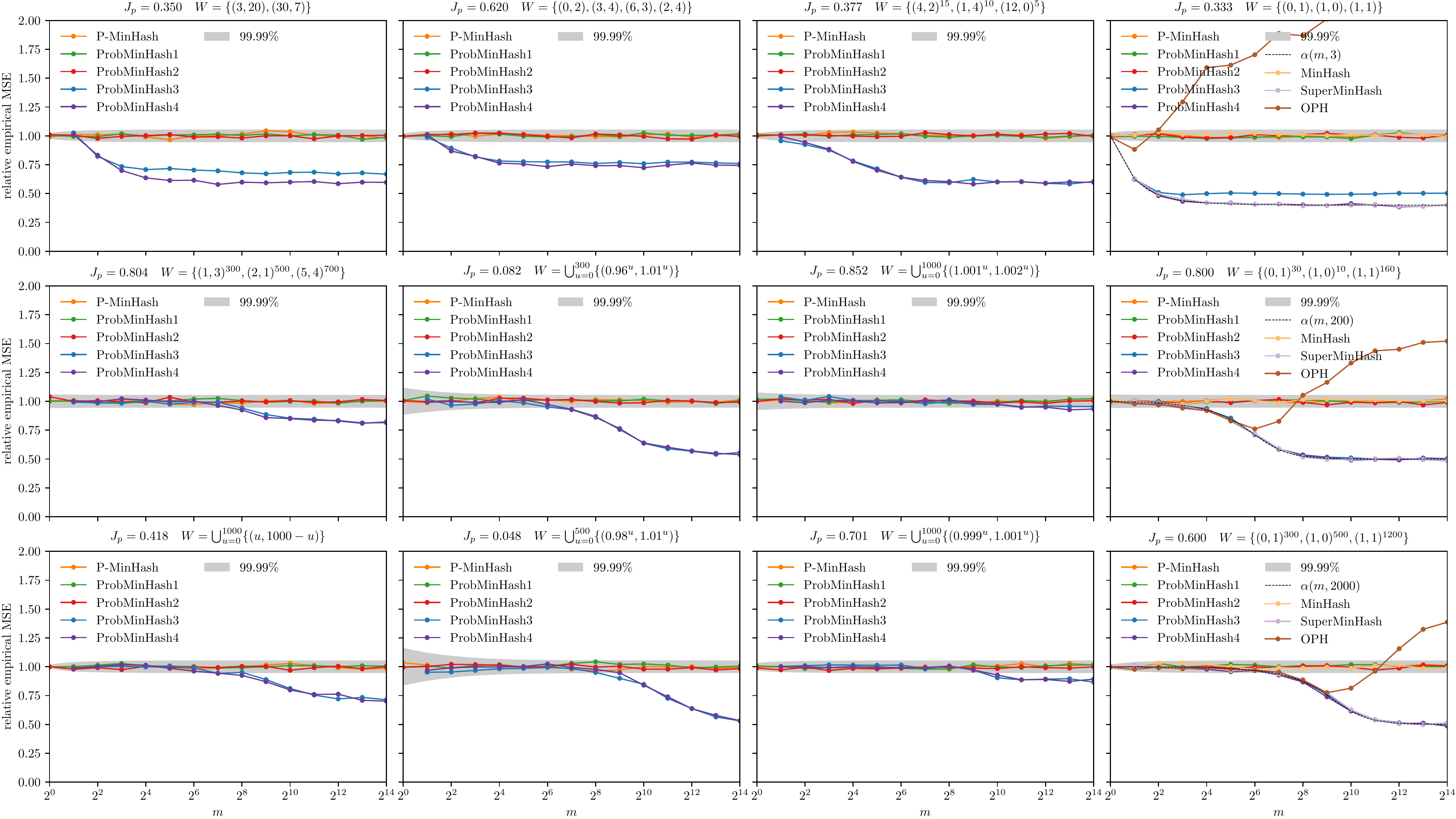}
\caption{The empirical \acf*{MSE} relative to $\symJaccardP (1-\symJaccardP)/\symHashSize$ over the signature size $\symHashSize$ for different multisets of weight pairs $\symTestWeightSet$. Each data point is calculated from \numprint{10000} pairs of randomly generated sets with weight functions satisfying $\symTestWeightSet = \bigcup_{\symInputElement:\symWeight_\symSetA(\symInputElement)>0 \vee \symWeight_\symSetB(\symInputElement)>0} \{(\symWeight_\symSetA(\symInputElement),\symWeight_\symSetB(\symInputElement))\}$. The gray band covers the middle 99.99\% of all $\symZScore$-scores, if the number of equal signature components is binomially distributed with success probability $\symJaccardP$.}
\label{fig:error}
\end{figure*}

All our experiments were performed with synthetic data. The reason for this is that realistic data sets usually do not contain enough different pairs of sets that have exactly the same predefined similarity. However, this is fundamental to verify the theoretically predicted distribution of estimation errors. We made the experience that testing locality-sensitive hash functions with realistic data sets is not very reliable. For example, if the algorithms presented in \cite{Yang2017, Raff2018, Yang2019a} had been checked with synthetic data using our test setup, it would have been easy to find that they are not suitable for the claimed metrics. More such examples can be found in \cite{Ertl2018}. Another advantage of synthetic data is that it facilitates the reproduction of our results. Although the test data consists of sets of randomly generated 64-bit integers, they are also representative for real-world situations. Realistic elements can always be reduced by hashing to 64-bit integers first, and with a good hash function they do not differ from the random numbers of our synthetic data sets.

For comparison, all experiments were also performed with P-MinHash \cite{Moulton2018} which is the state of the art for the probability Jaccard similarity $\symJaccardP$. Other algorithms for weighted sets like \ac{ICWS} \cite{Ioffe2010} or BagMinHash \cite{Ertl2018} were not considered, because they serve to calculate signatures for different Jaccard similarities $\symJaccardW$ or $\symJaccardN$, as mentioned in \cref{sec:introduction}. 
Since these algorithms are also more complex, they tend to be much slower than the algorithms for $\symJaccardP$ anyway.
For our examples with binary weights $\symWeight(\symInputElement)\in\{0,1\}$, where $\symJaccardP$ corresponds to the conventional Jaccard similarity $\symJaccard$, we also compared our ProbMinHash algorithms to MinHash, \ac{OPH} with optimal densification \cite{Shrivastava2017}, and SuperMinHash \cite{Ertl2017}. The latter two represent the state of the art of one-pass algorithms for the conventional Jaccard similarity $\symJaccard$.

\subsection{Verification}

\begin{figure*}[t]
\centering
\includegraphics[width=\textwidth]{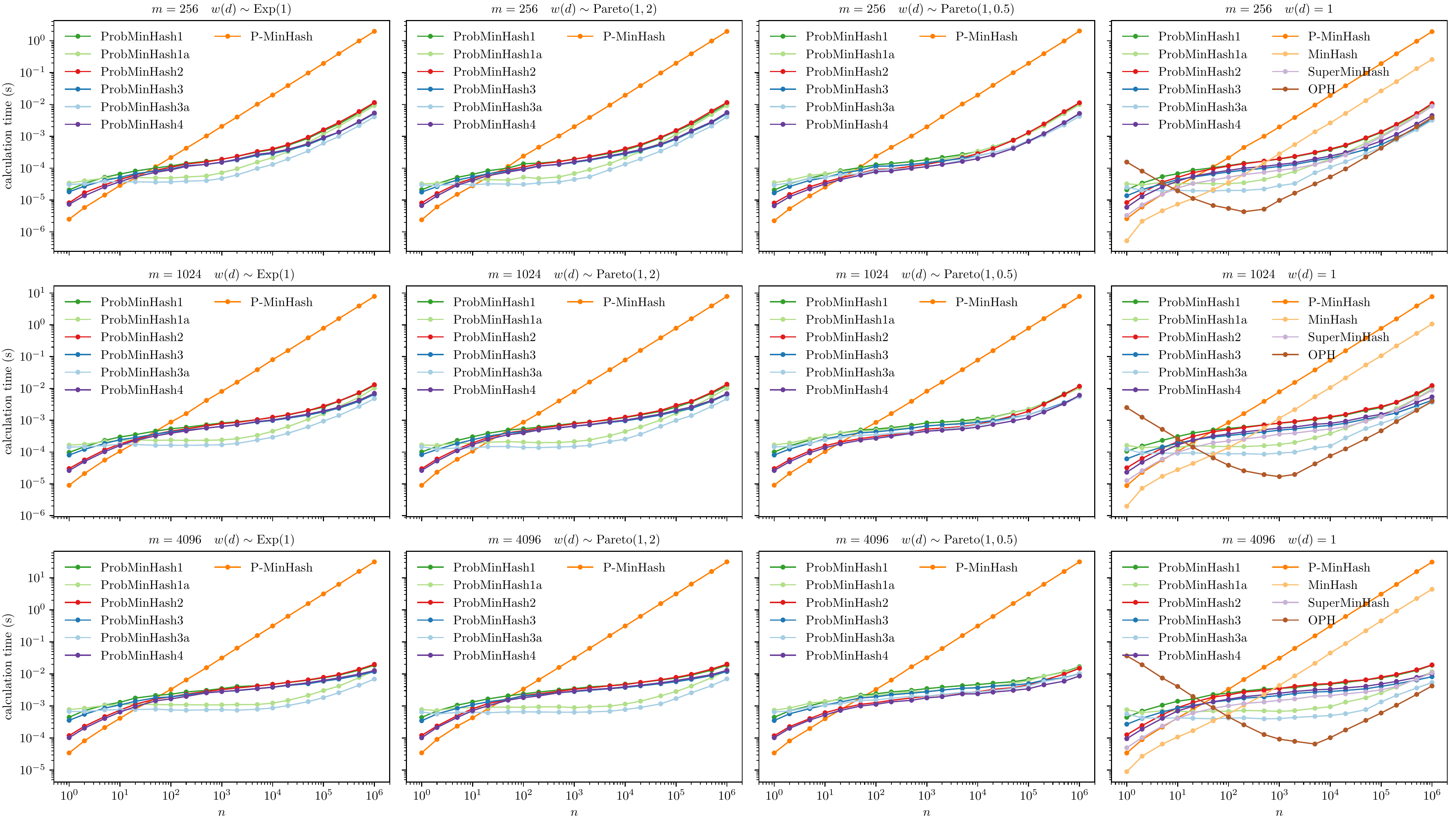}
\caption{The average calculation time for different signature sizes $\symHashSize$ and for different assumed distributions of $\symWeight(\symInputElement)$ over the input size $\symInputSize$.}
\label{fig:speed}
\end{figure*}

To verify our proposed algorithms we applied them to 12 different examples. Each example is characterized by a multiset $\symTestWeightSet$ of weight value pairs. Each pair represents the weights of some random element in two different sets, respectively. Given $\symTestWeightSet$, arbitrary many pairs of weighted sets $\symSetA$ and $\symSetB$ can be generated whose weight functions satisfy
$\symTestWeightSet = \bigcup_{\symInputElement:\symWeight_\symSetA(\symInputElement)>0 \vee \symWeight_\symSetB(\symInputElement)>0} \{(\symWeight_\symSetA(\symInputElement),\symWeight_\symSetB(\symInputElement))\}$. This is done by drawing a random element for each weight value pair in $\symTestWeightSet$, which is then added to sets $\symSetA$ and $\symSetB$ together with the corresponding weights, respectively. The resulting pairs of weighted sets will always have the same probability Jaccard similarity $\symJaccardP$ by definition as it is uniquely defined by $\symTestWeightSet$.

\numprint{10000} different pairs of such weighted random sets have been generated for each example. After computing the corresponding signatures with sizes $\symHashSize\in\{1,2,4,8,\ldots,2^{14}\}$, the similarity was estimated using \eqref{equ:jaccard_estimator} and the empirical \ac{MSE} with respect to the true $\symJaccardP$ was calculated. Since the expected empirical \acp{MSE} for P-MinHash and for the statistically equivalent algorithms ProbMinHash1 and ProbMinHash2 are equal to $\symJaccardP(1-\symJaccardP)/\symHashSize$, we considered the corresponding relative empirical \ac{MSE} which was finally plotted over the signature size $\symHashSize$ for all 12 examples in \cref{fig:error}. The regions covering the middle $99.99\%$ of the expected $\symZScore$-scores are also shown to indicate realistic deviations from the expected empirical \ac{MSE}. The $\symZScore$-score is obtained from the empirical \ac{MSE} by normalization. The expectation and the variance of the empirical \ac{MSE} are given by  $\symJaccardP(1-\symJaccardP)/\symHashSize$ and $\frac{\symJaccardP^2(1-\symJaccardP)^2}{\symHashSize^2\symNumExamples}(2-\frac{6}{\symHashSize})+\frac{\symJaccardP(1-\symJaccardP)}{\symHashSize^3\symNumExamples}$, respectively \cite{lonza}, where $\symNumExamples=\numprint{10000}$ is the sample size. 

The empirical \ac{MSE} for P-MinHash, ProbMinHash1, and ProbMinHash2 agree with the theoretical prediction, as the observed relative error is actually close to 1 and within the expected variation range. However, for ProbMinHash3 and ProbMinHash4 we observed that the error is significantly smaller, especially if the signature size $\symHashSize$ exceeds the input size $\symInputSize$. As both algorithms are not defined for $\symHashSize=1$ the corresponding points are missing in \cref{fig:error}. Dependent on the example, the correlated generation of points is able to reduce the empirical \ac{MSE} by up to a factor of two. We also observed that the empirical \ac{MSE} of ProbMinHash4 is slightly smaller than that of ProbMinHash3 for tiny sets as can be seen for the examples with  $\symTestWeightSet=\{(3,20),(30,7)\}$ and $\symTestWeightSet=\{(0,1),(1,0),(1,1)\}$.

The last column of \cref{fig:error} shows the results for three examples with binary weights for which we applied the unweighted variants of the ProbMinHash algorithms as described in \cref{sec:conventional_jaccard}. Furthermore, we also considered MinHash, \ac{OPH} with optimal densification \cite{Shrivastava2017}, and SuperMinHash \cite{Ertl2017}. Due to the independent signature components, the relative error of MinHash is as expected equal to 1. The theoretical relative \ac{MSE} of SuperMinHash is given by \eqref{equ:improvement_factor} and also shown in \cref{fig:error}. Since ProbMinHash4 corresponds to SuperMinHash in the unweighted case, the empirical \acp{MSE} of both were in perfect agreement with the theoretical prediction. The relative error of \ac{OPH} is larger than that of SuperMinHash and may even be much larger than that of MinHash for small $\symInputSize$ compared to $\symHashSize$. ProbMinHash1a and ProbMinHash3a have also been covered by our experiments. However, the logical equivalence to ProbMinHash1 and ProbMinHash3, respectively, led to identical results which were therefore omitted in \cref{fig:error}.

\subsection{Performance}

\begin{figure*}[t]
\centering
\includegraphics[width=\textwidth]{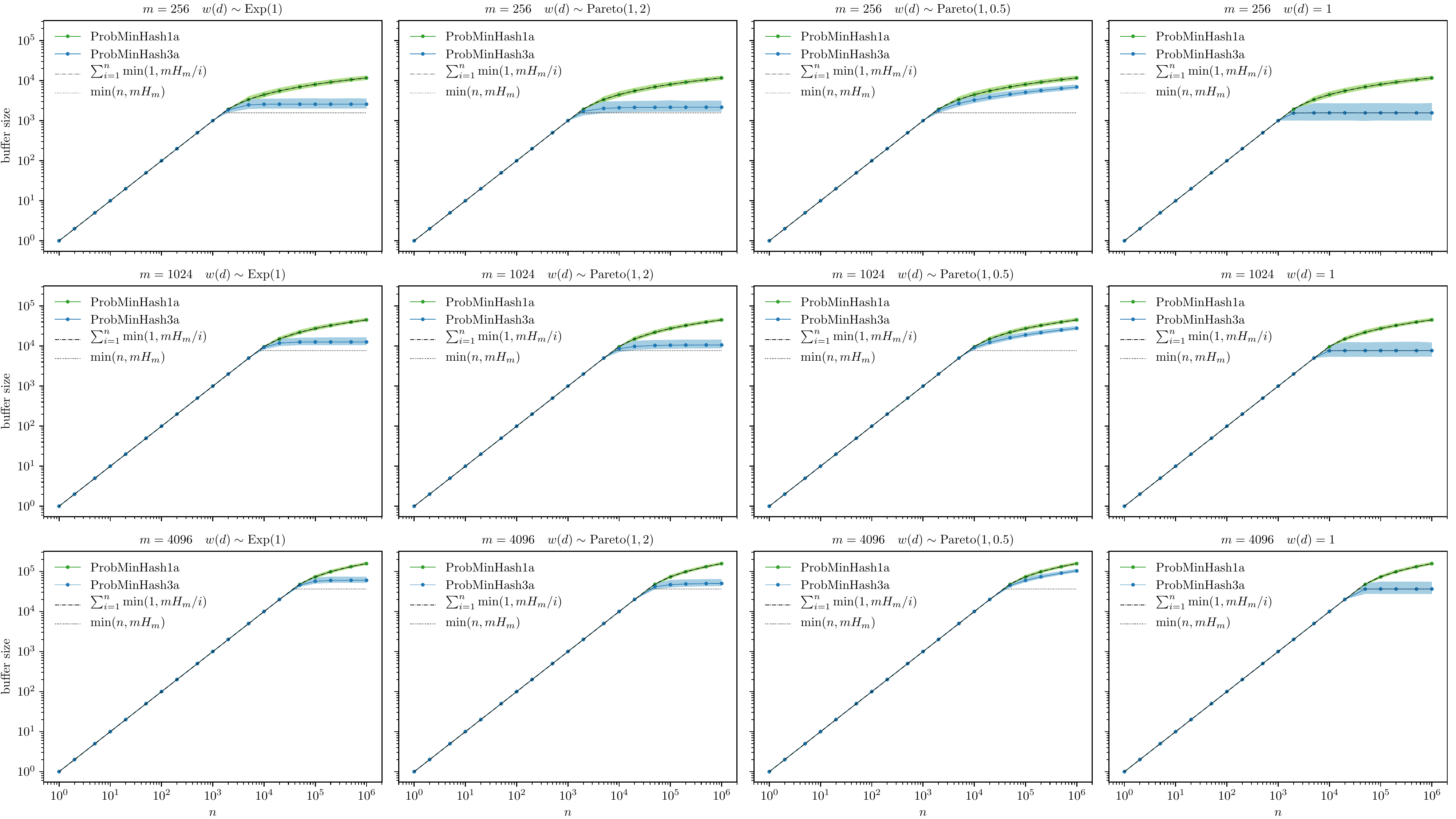}
\caption{The average maximum buffer size as a function of the number of input elements $\symInputSize$ for ProbMinHash1a and ProbMinHash3a. The shown bands cover the middle 99\% of all observed values.} 
\label{fig:buffer}
\end{figure*}

The performance of the presented algorithms was analyzed on a Dell Precision 5530 notebook with an Intel Core i9-8950HK processor and 32\,GB of memory. The average calculation time for a single signature was measured for signature sizes 256, 1024, and 4096, and different assumed weight distributions. 
\cref{fig:speed} shows the results as log-log plots over the input size $\symInputSize$ from $1$ to $10^6$. 
For each data point we first generated 100 randomly weighted sets and kept them in main memory to avoid distortion of the measurement. Only then the time for calculating the corresponding signatures was taken.

All ProbMinHash algorithms are significantly faster than the linearly scaling P-MinHash algorithm for large $\symInputSize$ with break-even points between 10 and 100. The maximum speedup factor is in the order of $\symHashSize$, because only the first point needs to be calculated for most elements compared to the $\symHashSize$ hash values per element in case of P-MinHash. The maximum speedup of ProbMinHash3, ProbMinHash3a, and ProbMinHash4 tends to be slightly higher, because the random value generation is less costly for the truncated than for the regular exponential distribution. P-MinHash is faster for very small $\symInputSize$. In the extreme case $\symInputSize=1$, all ProbMinHash variants which sample with replacement have a time complexity of $\symBigO(\symHashSize\log\symHashSize)$ and are therefore significantly slower. ProbMinHash2 and ProbMinHash4 do slightly better, because they sample without replacement and have like P-MinHash a complexity of $\symBigO(\symHashSize)$. Nevertheless they are still about 3 times slower than P-MinHash.

The last column in \cref{fig:speed} presents the results of examples with binary weights for which we also measured the performance of MinHash, \ac{OPH} with optimal densification \cite{Shrivastava2017}, and SuperMinHash \cite{Ertl2017}. P-MinHash is about an order of magnitude slower than MinHash, because the random value generation is more expensive for the exponential distribution than for the uniform distribution. 
Comparison of the ProbMinHash algorithms with MinHash gives break-even points between $\symInputSize=50$ and $\symInputSize=1000$.
The specialized variants of ProbMinHash3, ProbMinHash3a, and ProbMinHash4 are faster than SuperMinHash for large $\symInputSize$, which comes from the more adaptive stop limit as discussed in \cref{sec:conventional_jaccard}. \ac{OPH} has by far the best performance for input sizes $\symInputSize$ which are in the order of $\symHashSize$, but is quite slow for small $\symInputSize$. 

Interleaved processing as described in \cref{sec:interleaved} is not always faster for intermediate input sizes $\symInputSize$ as observed in the third column of \cref{fig:speed} when comparing ProbMinHash1a and ProbMinHash3a with ProbMinHash1 and ProbMinHash3, respectively. This probably depends on whether the expected weight is finite, as a significant speedup was observed for the $\symPareto(1,2)$-distribution with a mean of 2 and not for the $\symPareto(1,0.5)$-distribution with an infinite mean. Since an infinite average weight is not common in practice, we almost always expect an improvement with interleaved processing.

We also analyzed the additional space requirements of ProbMinHash1a and ProbMinHash3a. \cref{fig:buffer} shows the average and the middle 99\% of all observed buffer sizes when calculating the signatures of \numprint{10000} randomly generated weighted sets of size $\symInputSize$. The results for ProbMinHash1a perfectly match the theoretical considerations in \cref{sec:interleaved}. Also for ProbMinHash3a our prediction is correct, that the average buffer size for weight distributions with a power-tail with an index less than 1 reaches a plateau. In the unweighted case shown in the last column of \cref{fig:buffer} we could even confirm the predicted level of this plateau.

\subsection{Practical Considerations}
The optimal choice among the presented algorithms depends on the concrete application. The primary question to be answered is whether signature components should be independent or not. For example, if P-MinHash or MinHash should be replaced without changing the previous statistical behavior, or if it is desirable that signature components are statistically independent, as is the case for locality-sensitive hashing, we recommend using either ProbMinHash1a or ProbMinHash2. ProbMinHash1 is less relevant, because it is slower compared to ProbMinHash2 for very small input sizes $\symInputSize$ and does not have a worst case runtime complexity of $\symBigO(\symInputSize \symHashSize)$. ProbMinHash1a, which is logically equivalent to ProbMinHash1, might be more attractive as it is faster for medium $\symInputSize$, provided one is willing to sacrifice more memory.

On the other hand, if the signature is used for similarity estimation, we recommend ProbMinHash3a or ProbMinHash4. Both are based on correlated generation and therefore allow more accurate similarity estimations for $\symInputSize$ that are not much greater than the signature size $\symHashSize$. Our experiments have shown that ProbMinHash4 gives a better error reduction and is also faster for small $\symInputSize$ than ProbMinHash3, so that the latter is rather the second choice. However, if speed is most important, the logically equivalent ProbMinHash3a algorithm is recommended. It is the fastest variant for most input sizes $\symHashSize$ at the expense of additional memory. 

Since there are more options in the case of the conventional Jaccard similarity $\symJaccard$ and no algorithm completely outperforms all others, the optimal choice depends on the expected distribution of $\symInputSize$ and also on the requirements regarding the estimation error for input sizes $\symInputSize$ in the range of $\symHashSize$ or smaller. However, MinHash, ProbMinHash1a, or ProbMinHash2 should be used when independence of signature components is important.

\section{Outlook}
\begin{figure}[t]
\centering
\includegraphics[width=\columnwidth]{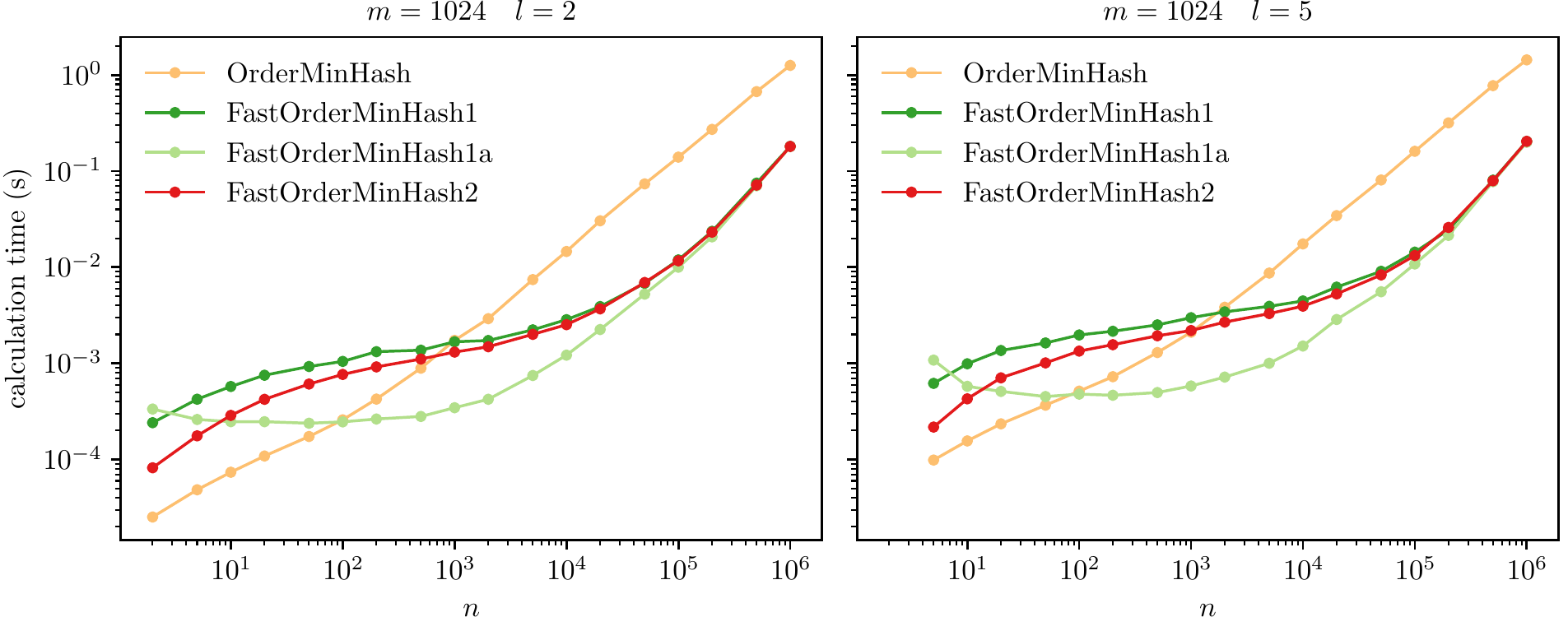}
\caption{The average calculation time of signatures for the edit similarity.}
\label{fig:order}
\end{figure}

Some other metrics might also benefit from the presented ideas. For example, there is the Lempel-Ziv Jaccard distance which is a generic similarity measure defined on the set of binary subsequences resulting during Lempel-Ziv compression \cite{Raff2019, Raff2017a}. Our fast algorithms could make it feasible to incorporate weights as proposed in \cite{Raff2017}. Another example is OrderMinHash, which was recently proposed as locality-sensitive hash algorithm for the edit similarity between sequences \cite{Marcais2019}. The signature calculation requires the smallest $\symIndexL$ hash values  for each component instead of just the smallest as with MinHash. We already implemented statistically equivalent algorithms called FastOrderMinHash1, FastOrderMinHash1a, and FastOrderMinHash2 based on ProbMinHash1, ProbMinHash1a, and ProbMinHash2, respectively. Our first results shown in \cref{fig:order} look very promising as the calculation time was reduced by an order of magnitude for longer sequences.

Apart from that, there are some open theoretical questions. We could not yet prove mathematically that the variance of ProbMinHash3 and ProbMinHash4 is never worse than that of P-MinHash. The only exception is the unweighted case of ProbMinHash4, which corresponds to the SuperMinHash algorithm and for which the variance is given by \eqref{equ:superminhash_variance}.
Futhermore, we assumed that elements are unordered with respect to their weights for our complexity analysis. It would be interesting to see how much an order affects the performance of the ProbMinHash algorithms. It is clear that it is advantageous to process elements in descending order with respect to their weights. The opposite is the case with ascending order.

\section{Conclusion}
We have introduced new locality-sensitive hash algorithms for the probability Jaccard similarity and, as a by-product, also for the conventional Jaccard similarity. These algorithms show largely a significantly better performance than other state-of-the-art methods. In addition, the calculated signatures partly allow a more precise similarity estimation. They can therefore improve the performance and accuracy of existing applications and also open up new applications. We therefore expect that our algorithms will soon be used in practice. 

\ifCLASSOPTIONcaptionsoff
  \newpage
\fi


\begin{IEEEbiography}[{\includegraphics[width=1in,height=1.25in,clip,keepaspectratio]{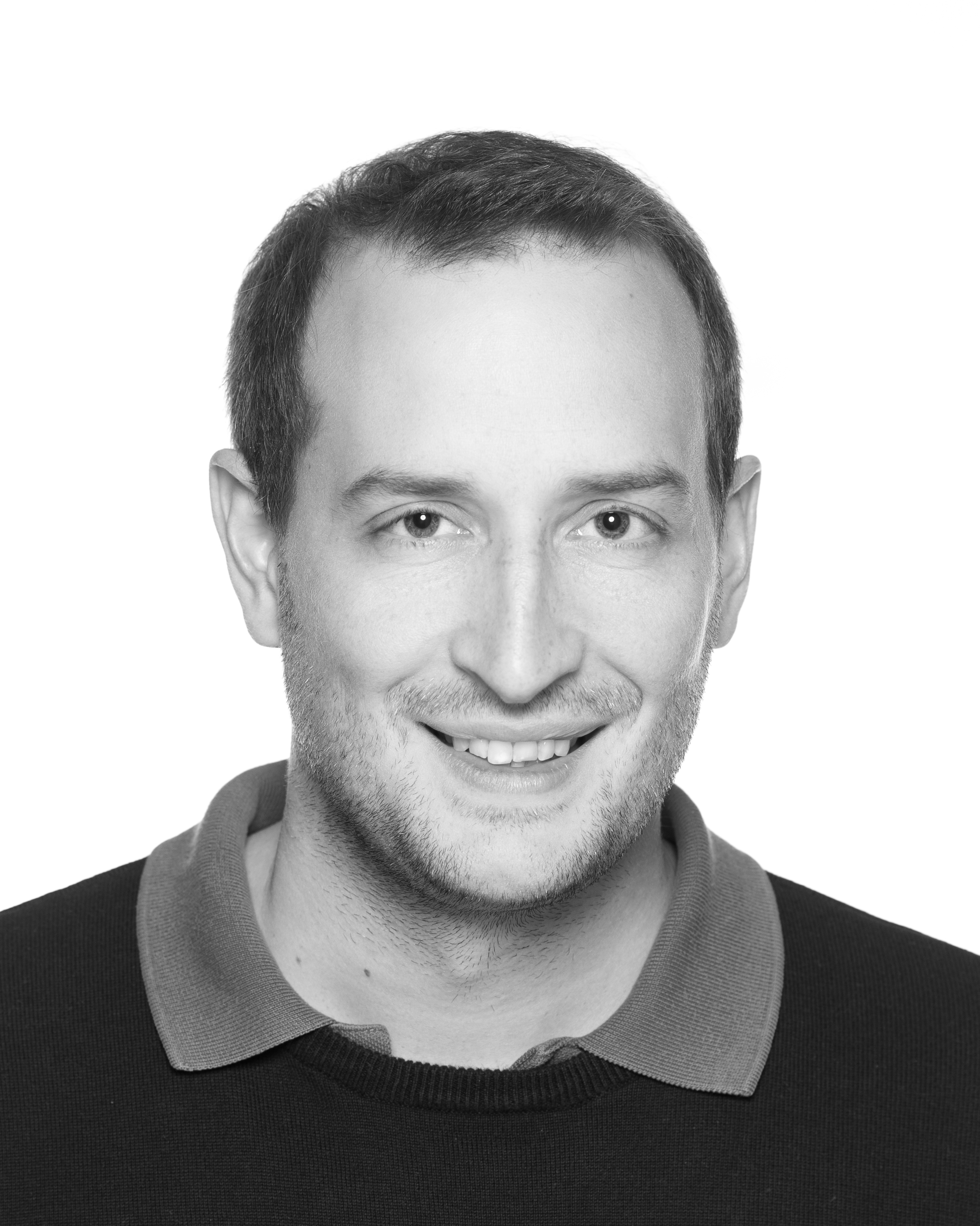}}]{Otmar Ertl}
is a Senior Software Mathematician at Dynatrace Research. He received the Ph.D. degree in technical
sciences from the Vienna University of Technology in 2010. His current research interests include probabilistic algorithms and data structures.
\end{IEEEbiography}
\end{document}